\pdfoutput=1
\documentclass[12pt]{article}
\usepackage{jheppub}
\usepackage{ifpdf}
\usepackage{graphicx}
\usepackage{microtype}
\usepackage{amsmath,amssymb}
\usepackage[utf8x]{inputenc}
\usepackage[T1]{fontenc}
\usepackage{ae}
\usepackage{aecompl}
\usepackage{titlesec}

\def\bfseries{\fontseries \bfdefault \selectfont \boldmath}

\usepackage{bm}
\usepackage{subfigure}

\usepackage[normalem]{ulem} 
\usepackage{xcolor,soul}

\definecolor{olive}{rgb}{0.3, 0.54, .1}

\titleformat*{\section}{\large\bfseries}
\titleformat*{\subsection}{\bfseries}
\titleformat*{\subsubsection}{\bfseries}

\newcommand{\bes}{\begin{eqnarray*}}
\newcommand{\ees}{\end{eqnarray*}}
\newcommand{\bel}[1]{\begin{eqnarray}\label{#1}}
\newcommand{\be}{\begin{eqnarray}}
\newcommand{\ee}{\end{eqnarray}}
\newcommand{\bea}{\begin{eqnarray}}
\newcommand{\eea}{\end{eqnarray}}

\newcommand{\nn}{\nonumber}

\newcommand{\mytitle}{ 
Critical behaviour of hydrodynamic series
}

\title{\boldmath\mytitle}

\author[1]{M.~Asadi}
\author[2,3,4]{H.~Soltanpanahi}
\author[1]{F.~Taghinavaz}

\affiliation[1]{ IPM, School of Particles and Accelerators, P.O. Box 19395-5531, Tehran, Iran}
 \affiliation[2]{Guangdong Provincial Key Laboratory of Nuclear Science, Institute of Quantum Matter, South China Normal University, Guangzhou 510006, China}
 \affiliation[3]{Guangdong-Hong Kong Joint Laboratory of Quantum Matter, Southern Nuclear Science Computing Center, South China Normal University, Guangzhou 510006, China}
 \affiliation[4]{Institute of Theoretical Physics, Jagiellonian University, S. Lojasiewicza 11, PL 30-348 Krakow, Poland}

\emailAdd{m\_asadi@ipm.ir}
\emailAdd{hesam.soltan@gmail.com}
\emailAdd{ftaghinavaz@ipm.ir}

\abstract{
We investigate the time-dependent perturbations of  strongly coupled $\mathcal{N}=4$ SYM theory at finite temperature and finite chemical potential with a second order phase transition. This theory is modelled by a top-down Einstein-Maxwell-dilaton description which is a consistent truncation of the dimensional reduction of type IIB string theory on AdS$_5\times$S$^5$. We focus on spin-1 and spin-2 sectors of perturbations and  compute the linearized hydrodynamic transport coefficients up to the third order in gradient expansion. We also determine the radius of convergence of the hydrodynamic mode in spin-1 sector and the lowest non-hydrodynamic modes in spin-2 sector. Analytically, we find that all the hydrodynamic quantities have the same critical exponent near the critical point $\theta=\frac{1}{2}$. Moreover, we establish a relation between symmetry enhancement of the underlying theory and vanishing the only third order hydrodynamic transport coefficient $\theta_1$, which appears in the  shear dispersion relation of a conformal theory on a flat background.
}

\begin{document}

\maketitle

\section{Introduction}\label{sec:intro}

The collisions of heavy ion at relativistic energies produce a hot and dense nuclear matter composed of deconfined quarks and gluons known as the strongly coupled quark-gluon plasma (QGP). 
The studies of the QGP at the Relativistic Heavy Ion Collider (RHIC) and the Large Hadron Collider (LHC) have led to the important results that explore a wide variety of QGP-related phenomena \cite{Rischke:2003mt, Shuryak:2003xe, Shuryak:2004cy}.
To describe in and out of equilibrium properties of this new phase at extreme conditions, Relativistic Hydrodynamics (RH) is a powerful tool \cite{Kovtun:2012rj, Romatschke:2017ejr, Florkowski:2017olj}.
The RH approach has great triumphs to explain the experimental events of the collider labs. Furthermore,
the condensed matter physics has also benefited from the RH applications \cite{Gooth:2017mbd}. All of these evidences show the "unreasonable effectiveness" of the RH irrespective of the energy scales \cite{Noronha-Hostler:2015wft}.

On the other hand, experimental observations imply that the QGP is a strongly interacting matter and  the perturbative calculations cease to apply \cite{Shuryak:2003xe}. Therefore, non-perturbative methods such as the  gauge/gravity duality  may shed some lights on various properties of the QGP. The well-known example of the AdS/CFT correspondence  states that the type IIB supergravity on the AdS$_5 \times$S$^5$  is dual to the 4$-$dimensional super Yang-Mills (SYM) theory living on the boundary of the  AdS$_5$ \cite{Maldacena:1997re, Witten:1998qj, Aharony:1999ti}. Indeed, this duality is a strong-weak duality which maps a strongly-coupled quantum gauge field theory to a weakly-coupled classical gravity in one higher dimension. One of the main consequences of the AdS/CFT application is the prediction of universal  ratio of shear viscosity to entropy density,  $\frac{\eta}{s}=\frac{1}{4\pi}$, which agrees very well with the experimental data \cite{Heinz:2013th, Policastro:2001yc}. 

The RH approach is an effective theory  for long  wave-length regime in which each conserved quantity can be expressed in terms of a gradient expansion \cite{Kovtun:2012rj}.  
 The coefficients of this series are the so called transport coefficients which contain the information of the underlying microscopic theory. 
One of the main concern regarding an infinite series expansions is the convergence features.
The  divergences which exist in the RH series in the real space \cite{Heller:2013fn} can be handled by using  the Borel-Pad\'e techniques \cite{Heller:2015dha, Heller:2016rtz,  Aniceto:2015mto, Aniceto:2018uik, Shokri:2020cxa}.
This provides  good information about the origin of divergent points and even the region in which the RH series  might be convergent. 
On the other hand, the convergence of the  gradient expansion in real space can be related to the radius of convergence of RH in momentum space \cite{Heller:2020uuy, Heller:2020jif}.
In the momentum space, the location of the singular points reflect  the existence of non-hydro modes  and the credit of the RH depends on the strength of hydro modes over non-hydro modes \cite{Romatschke:2015gic, Heller:2020hnq}.

Recently, the radius of convergence of hydrodynamics series have been investigated in various holographic cases. 
The radius of convergence of the shear-mode series in 4-dimensional AdS-Reissner-Nordstrom (AdS-RN) black brane has been studied in \cite{Withers:2018srf}. 
It has been extended to full range of the charge all the way to the extremal value both in 4 and 5-dimensional AdS-RN black branes in \cite{Jansen:2020hfd} (see also \cite{Abbasi:2020ykq}) and different types of pole collisions at different values of the charge has been found. Likewise, the analytic properties of dispersion relations in all sectors have been studied for $\mathcal{N}=4$ SYM theory at zero chemical potential dual to AdS$_5$ black branes in \cite{Grozdanov:2019kge, Grozdanov:2019uhi}.

In this paper,  we consider the 1-R charge black hole (1RCBH)\footnote{Throughout this paper, by 1RCBH we mean black hole with flat horizon.} model which is an analytical top-down string theory construction \cite{Gubser:1998jb, Behrndt:1998jd, Kraus:1998hv, Cai:1998ji, Cvetic:1999ne, Cvetic:1999rb}. It is obtained from 5-dimensional maximally supersymmetric gauged supergravity and is holographically dual to a 4-dimensional strongly coupled $\mathcal{N}=4$  SYM theory   at finite chemical potential under a $U(1)$ subgroup of the global $SU(4)$ symmetry of R-charges.
The interesting feature of this model is the existence of a second order phase transition at a certain point of the parameter space called  critical point. This phase transition belongs to the model B dynamical universality class \cite{Hohenberg:1977ym} and due to the large $N_c$ approximation, this phase transition shares same critical exponents as a mean-field  model \cite{Buchel:2010gd, Natsuume:2010bs}.

Different aspects of this background have been investigated. 
Indeed, it was shown that various quantities such as R-charge conductivity \citep{DeWolfe:2011ts}, complexity \citep{Ebrahim:2018uky},  mutual information \cite{Ebrahim:2020qif} and
entanglement of purification \cite{Amrahi:2020jqg, Amrahi:2021lgh} remain finite,  while their slopes diverge at the critical point with the same critical exponent $\theta=\frac{1}{2}$. 
The value of the dynamical critical exponent was also confirmed in \cite{Finazzo:2016psx} and \cite{Ebrahim:2017gvk} by studying the non-hydrodynamical quasi normal modes (QNMs) of the external fields and quantum quench in this background, respectively. 

Here, we would like to  derive the transport coefficients for the dual theory  of this background and  benefit from the standard  recipe \cite{Son:2002sd, Policastro:2002se, Kovtun:2004de, Son:2007vk}. We perform this calculation up to the third order of gradient expansion.  In spin-2 sector we compute the relaxation time $\tau_\pi$ and $\kappa$  in the second order, and $\lambda_{17}^{(3)}$ and $\lambda_1^{(3)} - \lambda_{16}^{(3)}$ in the third order of expansion from the Kubo formula.
Furthermore, in the spin-1 sector we  obtain another third order transport coefficient,  $\theta_1\equiv-(\lambda_1^{(3)}+\lambda_2^{(3)}+\lambda_4^{(3)})$,  by investigating the dispersion relation of shear hydro mode.

According to  the holographic dictionary,  the hydrodynamic excitations  correspond to the QNMs associated with poles of retarded Green's function of the conserved currents. Additionally, there is an infinite series of non-hydrodynamic QNMs in each sector resembling of the  Christmas trees \cite{Kovtun:2005ev}. 
 We investigate the radius of convergence of hydrodynamic series in the spin-1 sector for the whole range of the parameter space by computing the corresponding QNMs. In particular, we want to explore whether there is a critical behaviour and if so what is the associated  critical exponent. We show that, depend on the value of the ratio $\frac{\mu}{T}$, different forms of mode collision can  happen. Near the critical point there is a level-crossing between the hydro-mode and lowest non-hydro mode which is originally from the gauge field perturbation. While near the zero chemical potential  the radius of convergence is determined by a level-crossing between the hydro-mode and the lowest non-hydro mode from the gravity perturbation.

To study the linearized equations of motion,  enormous simplification will occur if we use the master equations formalism \cite{Kodama:2003jz, Kodama:2003kk, Jansen:2019wag} which has two great advantages. 
First, we can investigate the set of equations analytically to compute higher order transport coefficients and also to find the radius of convergence of the shear hydrodynamic series.
Second, it speeds up significantly the numerical computations of QNM in the spin-1 sector. 

The organization of this paper is as follows. 
In section \ref{sec:hydro} we review the building blocks of hydrodynamics and gradient expansion and introduce the transport coefficients  up to the third order in gradient expansion for the conformal hydrodynamics. 
In section \ref{sec:GreensF} we give the preliminary ingredients about the Green's function and how to derive the transport coefficients from the two-point functions in the AdS/CFT formalism via the Kubo formula.
Section \ref{sec:1RCBH} is devoted to review the thermodynamics of the 1RCBH model and fix our notation.  
In section \ref{sec:hydro-1RCBH} we  study the hydrodynamics of the 1RCBH model and illustrate the details for perturbations in spin-2 and spin-1 sectors. The second and third order transport coefficients as well as the dispersion relation of the shear mode are derived. In section \ref{sec:QNM} we compute the QNM frequencies in spin-2 and  spin-1 sectors for complex momenta. We find the radius of convergence of the hydrodynamic series in shear channel using both analytical and numerical approaches. Likewise, in spin-2 sector  the radius of convergence of lowest non-hydro modes is determined.
In section \ref{sec:critical} we investigate the behaviour of the transport coefficients and the hydrodynamic radius of convergence near the critical point of the phase transition and we show that all of them exhibit the same critical exponent $\theta=\tfrac{1}{2}$. 
We conclude with a summary and an outlook to further directions in section \ref{sec:conclusion}.

\section{Hydrodynamics as derivative expansion}\label{sec:hydro}

In this section we will review the basic principles of the relativistic hydrodynamics of the boundary theory.
The existence of the hydrodynamic equations is due to the conservation laws which are related to the continuous symmetries of the underlying theory. These symmetries yield the conserved quantities whose fluctuations are long-lived and long distance propagating modes, i.e. $\omega\rightarrow0$ as ${k\rightarrow0}$.
 If one considers a relativistic hydrodynamic including a local $U(1)$ symmetry, then the hydrodynamic equations can be read as\footnote{We  use the capital Latin letters ($M, N$, \dots) for the bulk coordinates and Greek letters ($\mu, \nu$, \dots) for the boundary coordinates. We also adopt the convention $\hbar = c = 1$.}  
\begin{align}\label{eq6}
	\nabla_\mu T^{\mu \nu}= F^{\nu \mu} J_\mu, \qquad \nabla_\mu J^{\mu } = 0,
\end{align} 
where $T^{\mu\nu}$ is the energy-momentum tensor corresponding to the space-time symmetries and $J^\mu$ is  a conserved current  corresponding to a local $U(1)$ symmetry of the underlying  theory. In order to solve the Equations \eqref{eq6}, one may use the fact  that $T^{\mu \nu}$ and $J^{\mu }$ can be expressed as functions of the local temperature $T(x^\nu)$, the local four velocity $u^{\mu}(x^\nu)$ and the local chemical potential $\mu(x^\nu)$ which are known as the hydrodynamic variables. 

 Usually, the constitutive relations for the energy-momentum tensor and the current density are written in the Landau frame  as
 \begin{eqnarray}\label{eq:constitutive}
T^{\mu\nu}&=&\epsilon u^{\mu}u^{\nu}+p\Delta^{\mu\nu}+\Pi^{\mu\nu},\nn\\
J^\mu&=& n u^\mu + j^\mu,
\end{eqnarray}  
where $\left(\epsilon, p, n\right)$ are the equilibrium energy density, pressure and charge density, respectively. In flat space-time the operator $\Delta^{\mu \nu}\equiv \eta^{\mu\nu}+u^\mu u^\nu$ is a projector perpendicular to the fluid velocity $u^\mu$ and the flat metric $\eta^{\mu\nu}=\textbf{diag}(-1,1,1,1)$. The $\Pi^{\mu \nu}$ and $j^\mu$ tensors are dissipative contributions which can be expressed in terms of the derivatives of the hydrodynamic variables \cite{Kovtun:2012rj, Romatschke:2017ejr, Florkowski:2017olj}. 
If the underlying theory possess the conformal symmetry, then the gradient expansion of the  dissipative parts  can be written as
 \cite{Baier:2007ix, Grozdanov:2015kqa}
\begin{align}\label{eqpi}
    \Pi^{\mu \nu} &= -\eta \sigma^{\langle \mu \nu \rangle}+ \eta \tau_\pi \bigg(D \sigma^{\langle \mu \nu \rangle} + \frac{1}{3} \sigma^{\langle \mu \nu \rangle} \nabla \cdot u\bigg) + \kappa \bigg(R^{\langle \mu \nu \rangle} - 2 u_\alpha R^{\alpha \langle \mu \nu\rangle \beta} u_\beta\bigg)\nonumber\\
&+{\lambda}_1^{(2)} \sigma^{\langle\mu}\,_{\alpha}\sigma^{\nu\rangle \alpha} + {\lambda}_2^{(2)} \sigma^{\langle\mu}\,_{\alpha}\Omega^{\nu\rangle\alpha} + \lambda_3^{(2)} \Omega^{\langle\mu}\,_{\alpha}\Omega^{\nu\rangle \alpha} + \sum_{i=1}^{20}\lambda_i^{(3)} \mathcal{O}_i^{(3) \mu \nu}+\mathcal{O}(\partial^4),\nn\\
j^\mu &= - \sigma T \Delta^{\mu \nu} \partial_\nu(\frac{\mu}{T}) + \mathcal{O}(\partial^2).
\end{align}
The structure of the operators are chosen such that they transform homogeneously under the Weyl transformations \cite{Baier:2007ix}. Among the transport coefficients appeared in \eqref{eqpi}, some of them are more familiar such as the shear viscosity, $\eta$, the relaxation time, $\tau_\pi$ and the conductivity, $\sigma$, while other ones are less  recognized, like the non-linearized second order transport coefficients $\lambda_i^{(2)}$ and the third order transport coefficients $\lambda_i^{(3)}$. Due to the Landau matching condition, dissipative terms are transverse to the velocity profile $u_\mu \Pi^{\mu \nu} = u_\mu j^\mu= 0$ and because of the conformal symmetry we have $g_{\mu \nu} \Pi^{\mu \nu} =0$. The shear stress tensor is defined as $\sigma^{\langle \mu \nu \rangle }\equiv \nabla^{\langle \mu} u^{\nu \rangle}$ and for a given second rank tensor we use the following notation
\begin{align}
    \mathcal{A}^{\langle \mu\nu\rangle} \equiv \frac{1}{2}\Delta^{\mu \alpha}\Delta^{\nu \beta}(\mathcal{A}_{\alpha\beta}+\mathcal{A}_{\beta\alpha})-\frac{1}{3}\Delta^{\mu \nu}\Delta^{\alpha \beta} \mathcal{A}_{\alpha\beta}.\nn
\end{align}
In addition, the vorticity field $\Omega^{\mu \nu}$ and the convective derivative $D$ are expressed as
\begin{align} 
& \Omega^{\mu \nu} = \frac{1}{2} \Delta^{\mu \alpha} \Delta^{\nu \beta} \bigg(\nabla_\alpha u_\beta - \nabla_\beta u_\alpha\bigg) ,\nonumber\\
&D\equiv u^{\mu} \nabla_{\mu}.\nn
\end{align}
The third order operators $\mathcal{O}_i^{(3)}$  in Equation \eqref{eqpi} are complicated Weyl-covariant tensors built out of the third order gradients whose detailed forms  can be found in Ref. \cite{Grozdanov:2015kqa}. The corresponding third order transport coefficients $\lambda_i^{(3)}$ are not uniquely defined but for a given underlying theory there exist at most $20$ of them.

The transport coefficients  $(\eta, \sigma)$ of the 1RCBH model  are calculated in the  \cite{DeWolfe:2011ts}, while the other ones  are still undetermined. It is shown that the shear viscosity and the conductivity exhibit a critical behavior close to the critical point of the second order phase transition \cite{DeWolfe:2011ts}. A natural question  is whether other hydrodynamic properties (e.g. the transport coefficients, the radius of convergence) of the underlying theory  would exhibit similar behaviour. 
In the following section we present the basic ingredients and methods to derive the transport coefficients for $\mathcal{N}=4$ SYM theory at finite temperature and finite chemical potential by using the gauge/gravity duality. In section \ref{sec:hydro-1RCBH} we will benefit from our results to derive the corresponding second and third order transport coefficients.

\section{Green's functions and transport coefficients}\label{sec:GreensF}

For a  strongly coupled field theory which has dual gravity interpretation, the AdS/CFT correspondence has a recipe to derive the transport coefficients \cite{Son:2002sd, Policastro:2002se, Kovtun:2004de, Son:2007vk}. In this section, our main focus is to review the building blocks of this recipe. This includes the formalism to obtain the retarded Green's functions, the variational approach and the Kubo formula. 

\subsection{Retarded Green's functions and the holographic principle}

To compute the $n$-point functions one needs to know the generating functional $Z(J)$ of the theory
\begin{align}
Z(J) = \int \mathcal{D}\Phi e^{-S[\Phi, J]}, \nn
 \end{align} 
where $J$ is the source term. The $n$-point functions are nothing but the  $n$'th functional derivatives of $Z[J]$ with respect to $J$ 
\begin{align}
 G_{(n)}(x_1, \ldots, x_n) = i^n \frac{\delta^n Z(J)}{\delta J(x_1) \ldots \delta J(x_n)}. \nn
 \end{align}
To our purpose, the two-point functions are of great importance. Consider a quantum field theory which has some gauge invariant operators, say $\hat{\mathcal{O}}_i$. Due to the causal structure of the relativistic hydrodynamics, we will focus only on the retarded two-point functions which are  defined in momentum space as
\begin{align}
G^R_{i j}(k) &= -i \int \, d^4x \,\, e^{-ik\cdot x}\, \theta(t) \langle \left[\hat{\mathcal{O}}_i(x), \hat{\mathcal{O}}_j(0)\right]\rangle,\nn
\end{align}   
where $k=(\omega , \overrightarrow{q})$. 
From the spectral representations, if one knows the retarded Green's function, then the two-point functions (advanced, symmetric and Feynman) will be easy to find \cite{Son:2002sd}.

In a holographic setup, there is a straightforward method to compute the retarded two-point functions of the boundary theory.
Suppose that  $J$ is a source for  an operator $\hat{\mathcal{O}}$ in the boundary theory
with an interaction Lagrangian density term  $J \,\hat{\mathcal{O}}$.
According to the Euclidean description of the AdS/CFT correspondence, we have the following relation
	\begin{align}\label{e9}
		\langle e^{\int_{\partial M} J \hat{\mathcal{O}}}\rangle = e^{-S_{cl}(\Phi)},
	\end{align} 
where the  left-hand side is the  expectation value of the generating functional for the  operator $\hat{\mathcal{O}}$ living on the boundary  $\partial M$, while the right-hand side is the exponent of  the classical on-shell action subject to the  boundary condition  $\Phi \big{|}_{\partial M}= J$. Note that the Equation \eqref{e9} is written in Euclidean signature while we are interested in Minkowski two-point functions. The procedure to make this connection is explained neatly in section 4 of Ref. \cite{Policastro:2002se} and we follow their recipe to compute $G^R_{i j}(k)$.

\subsection{Linear response theory}
The transport coefficients can be determined via the variational approach and Kubo formula \cite{Kovtun:2012rj}.
To quantify the hydrodynamics response, we use  the conserved currents, $T^{\mu\nu}(\mathbf{x})$ and $J^{\mu}(\mathbf{x})$, since they have microscopic definitions. 
One can use the variational approach to derive a specific two-point functions in the following steps \cite{Kovtun:2012rj}:
 \begin{enumerate}
 	\item Write down the constitutive relations and expand them up to the first order in the hydrodynamic fluctuations  $\left(\delta T, \delta \mu, \delta u^\mu\right)$ and the source fluctuations $\left(\delta A_\mu, \delta g_{\mu \nu}\right)$.
 	\item Use the Equation \eqref{eq6} to find the hydrodynamic fields $\left(\delta T, \delta \mu, \delta u^\mu\right)$ in terms of the source fields $\left(\delta A_\mu, \delta g_{\mu \nu}\right)$.
 	\item Plug the solutions into the constitutive relations \eqref{eq:constitutive} and make them on-shell.
 	\item Then, we use the on-shell currents to  define the retarded Green's functions as
 \begin{align}
	&G^R_{J^{\mu} J^{\nu}} \equiv - \frac{\delta (\sqrt{-g} \langle J^\mu\rangle) }{\delta A_\nu}\bigg|_{\delta g = \delta A=0}, \qquad
	G^R_{J^{\mu} T^{\alpha \beta}} \equiv - 2\frac{\delta (\sqrt{-g} \langle J^\mu\rangle)}{\delta h_{\alpha \beta}} \bigg|_{\delta g = \delta A=0},\nn\\
	&G^R_{T^{\mu \nu} J^{\alpha}} \equiv - \frac{\delta (\sqrt{-g} \langle T^{\mu \nu}\rangle)}{\delta A_\alpha}\bigg|_{\delta g = \delta A=0}, \qquad
	G^R_{T^{\mu \nu} T^{\alpha \beta}} \equiv - 2\frac{\delta (\sqrt{-g} \langle T^{\mu \nu}\rangle)}{\delta h_{\alpha \beta}}\bigg|_{\delta g = \delta A=0}.\nn
\end{align}
  \end{enumerate}
  
 Now we are in a position to obtain the transport coefficients. Each  Green's function corresponds to a specific set of transport coefficients \cite{Kovtun:2012rj}. 
We are interested in a 3+1 dimensional hydrodynamic theory with a SO(3) symmetry in the spacial directions. Using this symmetry we can set the momentum along the $z$ coordinate. To our purpose, we will focus on a so-called shear retarded Green's function $G^R_{xy, xy}\equiv 	G^R_{T^{xy} T^{xy}}$ which has the following gradient expansion up to the third order\cite{Baier:2007ix, Grozdanov:2015kqa}
 \begin{align}\label{eq37}
	G^R_{xy, xy}(\omega, q) =& p - i \eta \omega + \eta \tau_\pi \omega^2- \frac{\kappa}{2} \left(\omega^2 + q^2\right) \nn\\
	&- \frac{i}{2}\lambda_{17}^{(3)} \omega^3 + \frac{i}{2} (\lambda_1^{(3)} - \lambda_{16}^{(3)} - \lambda_{17}^{(3)}) \omega q^2.
\end{align} 
From this relation it is transparent that the transport coefficients can be found by taking special limits of $G^R_{xy, xy}$ which is the well-known Kubo formula \cite{Baier:2007ix, Kovtun:2012rj, Grozdanov:2015kqa}. 
Here is the explicit form of the Kubo formula for each transport coefficient in this channel,
\begin{align}\label{eq37b} 
&\eta=-\lim\limits_{\omega\to 0}\lim\limits_{q\to 0}\frac{d}{d\omega}Im\, G^R_{xy, xy},\nn\\
&\kappa=-2\lim\limits_{\omega\to 0}\lim\limits_{q\to 0}\frac{d}{dq^2}Re\, G^R_{xy, xy},\nn\\
&\eta\tau_\Pi-\frac{\kappa}{2}=\lim\limits_{\omega\to 0}\lim\limits_{q\to 0}\frac{d}{d\omega^2}Re\, G^R_{xy, xy},\\
&\lambda_{17}^{(3)} = -2\lim\limits_{\omega\to 0}\lim\limits_{q\to 0}\frac{d}{d\omega^3}Im\, G^R_{xy, xy},\nn\\
& \lambda_{1}^{(3)} - \lambda_{16}^{(3)} - \lambda_{17}^{(3)} = 2\lim\limits_{\omega\to 0}\lim\limits_{q\to 0}\frac{d^2}{d\omega dq^2}Im\, G^R_{xy, xy}.\nn
\end{align}
These relations are universal which can be applied to any conformal field theory, including one with  dual gravity description. For example, in the context of AdS/CFT correspondence and by using Equation \eqref{eq37b} one can find the first and second order transport coefficients of the strongly interacting $\mathcal{N}=4$ SYM \cite{,Baier:2007ix, Son:2002sd, Policastro:2002se, Kovtun:2004de, Son:2007vk}. Although, some difficulties may arise such as implementing the holographic renormalization to remove the divergent terms  \cite{Elvang:2016tzz, Skenderis:2002wp, Skenderis:2008dg}, but the calculation is straightforward. In section \ref{sec:hydro-1RCBH}, we will show  how to utilize the holographic machinery as well as the Equation \eqref{eq37b} to obtain the transport coefficients of a 4-dimensional CFT with second order phase transition which is dual to the 1RCBH backgrounds.

\section{Holographic model: 1RCBH}\label{sec:1RCBH}

In this section, we present a short review on the gravity setup which  mimics a second order phase transition of the boundary theory. The bulk theory is a top-down string theory construction which is a consistent truncation of the super-gravity on AdS$_5\times$S$^5$ geometry keeping only one scalar field and one gauge field coupled to the Einstein gravity. The thermal solution is an asymptotically AdS black brane geometry with nontrivial profile of the scalar and gauge fields, so-called 1RCBH \cite{Gubser:1998jb,Behrndt:1998jd,Kraus:1998hv,Cai:1998ji,Cvetic:1999ne,Cvetic:1999rb}. This geometry is dual to a four dimensional strongly coupled gauge theory $\mathcal{N}=4$ SYM at finite temperature and chemical potential.

\subsection{The background geometry}

The 1RCBH model is described by the following Einstein-Maxwel-dilaton (EMD) action
\begin{equation}\label{1R-action}
S_\textrm{bulk}=\frac{1}{16 \pi G_5} \int d^{5}x \sqrt{-g} \bigg{[}R-\frac{f(\phi)}{4} F_{M N} F^{M N} -  \frac{1}{2} \partial_{M} \phi \partial^{M} \phi-V(\phi) \bigg{]}.
\end{equation}
where $G_5$ is the five dimensional Newton's constant.
The self-interacting dilaton potential $V(\phi)$ and the Maxwell-dilaton coupling $f(\phi)$ are given by
\begin{align}
V(\phi)&=-\frac{1}{L^2}\big{(}8 e^{\frac{\phi}{\sqrt{6}}} + 4 e^{-\sqrt{\frac{2}{3}}\phi}\big{)},\nn\\
f(\phi)&=e^{-2 \sqrt{\frac{2}{3}}\phi} .\nn
\end{align} 
 where $L$ is the AdS$_5$ radius and without loose of generality we set $L=1$ now-on.
The corresponding  equations of motion are
\begin{align}\label{eq3}
		&\frac{1}{\sqrt{-g}} \partial_M\left(\sqrt{-g} g^{M N} \partial_N \phi\right) - \frac{f'(\phi)}{4} F_{MN}F^{MN} - V'(\phi)=0,\nonumber\\
		&\partial_M\left(\sqrt{-g} f(\phi) F^{MN}\right)=0,\\
		&R_{MN} - \frac{g_{MN}}{3} \left(V(\phi)-\frac{f(\phi)}{4} F^2 \right) - \frac{f(\phi)}{2} F_{M O} F_N^{~O} -\frac{1}{2}\partial_M \phi\partial_N\phi=0.\nn
	\end{align}
The charged static stationary black brane solutions are given by
\begin{align}
\label{metric}
ds^2 &= e^{2A(\tilde{r})} \left(-h(\tilde{r}) dt^2 + d\vec{x}^2\right) + \frac{e^{2B(\tilde{r})}}{\tilde{r}^4 h(\tilde{r})} d\tilde{r}^2,\nn\\
A(\tilde{r}) &=-\log \tilde{r} + \frac{1}{6} \log \left(1+ \widetilde{Q}^2 \tilde{r}^2\right),\nn\\
B(\tilde{r}) &=\log \tilde{r} - \frac{1}{3} \log \left(1+ \widetilde{Q}^2 \tilde{r}^2\right),  \nn\\
h(\tilde{r}) &= 1 - \frac{{\widetilde{M}}^2 \tilde{r}^4}{1+ \widetilde{Q}^2 \tilde{r}^2},\nn\\
\phi(\tilde{r}) &=-\sqrt{\frac{2}{3}} \log \left(1+\widetilde{Q}^2 \tilde{r}^2\right), \nn\\ 
\mathbf{A}(\tilde{r}) & =\widetilde{M} \widetilde{Q}\left(\frac{\tilde{r}_h^2}{1+\widetilde{Q}^2 \tilde{r}_h^2} - \frac{\tilde{r}^2}{1+\widetilde{Q}^2 \tilde{r}^2}\right)dt,
\end{align}
where  $\tilde{r}=0$ is the boundary of asymptotically AdS$_5$ background and $\widetilde{M}$ and $\widetilde{Q}$  are the  mass and charge of the black brane, respectively. The black hole horizon is given by the largest root of the blackening function, $h(\tilde{r})|_{\tilde{r}=\tilde{r}_h}=0$, 
\be
\tilde{r}_h=\sqrt{\frac{\sqrt{4 \widetilde{M}^2+\widetilde{Q}^4}+\widetilde{Q}^2}{2\widetilde{M}^2}},\nn
\ee
and the gauge field $\mathbf{A}$ has only a time component which vanishes at the horizon due to regularity conditions. 
We use the following reparametrization
	\begin{align}\label{eq:reparametrization}
		M\equiv r_h^2 \widetilde{M}, \qquad 
		Q\equiv r_h \widetilde{Q}, \qquad 
		r\equiv r_h \tilde{r},
	\end{align}
to introduce the dimensionless parameters. In this manner the horizon radius can be fixed to one which leads to a simple relation between the mass and charge, namely 
 \begin{equation}\label{eq:Mass}
M = \sqrt{1+Q^2}.
\end{equation}
Using the above dimensionless parameters  simplifies our calculation presented in the following sections. We will use this parameterization now-on.

 \subsection{Thermodynamic}

The Hawking temperature $T$  of the black brane is given by
\begin{align}
\label{temp}
T &=\frac{2+Q^2}{2 \pi \tilde{r}_h \sqrt{1+Q^2}},
\end{align}
which according to the gauge-gravity correspondence is equal to the temperature of the boundary theory. In addition, the chemical potential of the dual theory reads as
\begin{equation}\label{muo}
\mu = \lim\limits_{r\to 0} \mathbf{A}_t(r) = \frac{Q}{\tilde{r}_h\sqrt{1+Q^2}}.
\end{equation}
It is straightforward to see that once the charge parameter vanishes, we will recover the geometry of the AdS$_5$-Schwarzschild background as one may expect.

One can   characterize the 1RCBH model either by two non-negative parameters $(M,Q)$ from gravity point of view  or $(\mu ,T)$ from the boundary point of view.  Nevertheless, the reparameterization introduced in Equation \eqref{eq:reparametrization} can manifest an extra scaling symmetry in the system. Using Equations \eqref{temp} and \eqref{muo} one may make it more transparent  to find the charge as a ratio of the boundary theory parameters,
\begin{align}\label{eq7}
		Q  = \sqrt{2} \frac{1 \pm \sqrt{1-(\frac{\mu/T}{\pi/\sqrt{2}})^2}}{\left({ \frac{\mu/T}{\pi/\sqrt{2}}}\right)} .
	\end{align}
Since $Q$ is real the above equation indicates that $\frac{\mu}{T} \in [0, {\pi/ \sqrt{2}}]$. Moreover, Equation \eqref{eq7} shows that for a given value of $\frac{\mu}{T} \in \left[0, {\pi/ \sqrt{2}}\right)$ there are two distinct solutions corresponding to values of $Q$, while  $Q=\sqrt{2}~  (\mu=\pi T/ \sqrt{2})  $ is the merging point of the two branches. We will show that  thermodynamic quantities of the 1RCBH background diverge at the merging point declaring that this is the critical point of a second order phase transition point.  
In Ref. \cite{DeWolfe:2011ts} it was shown that  the solutions with $-/+$ sign in Equation. \eqref{eq7} are thermodynamically stable/unstable black branes. In this paper we are only interested in stable geometries and choose the branch corresponding to the $-$ sign in Equation. \eqref{eq7}. It turns out that introducing a new variable $y$ as 
\begin{align}
		y^2+\frac{2}{\pi^2}\left(\frac{\mu}{T}\right)^2 =1 ,\nn
		\qquad y\in [0, 1],
	\end{align}
both simplifies our equations  and  makes the investigations close to the critical point more intelligible. Accordingly, one may express the charge $Q$, the Hawking temperature $T$ and the chemical potential $\mu$ of the black brane in terms of this new dimensionless parameter
\begin{align}\label{eq9}
	Q = \sqrt{2 \frac{1-y}{1+y}},\qquad T= \frac{2}  {\pi \tilde{r}_h\sqrt{(3-y) (1+y)}},\qquad \mu=\frac{\pi  T \sqrt{1-y^2}}{\sqrt{2}}.
\end{align}
It is easy to see that the critical point $(Q=\sqrt{2})$ and the AdS$_5$-Schwarzschild background  $(Q=0)$ correspond to $y=0$ and $y=1$, respectively.

In the context of the AdS/CFT correspondence, we have $\frac{L^3}{\kappa_5^2}=\frac{N_c^2}{4\pi^2}$ where $\kappa_{5}$ is the five dimensional gravitational constant given by $\kappa_{5}^{2}=8\pi G_5$ . 
By using the Equation \eqref{eq9} the Bekenstein entropy density for the black brane geometry    \eqref{metric} can be computed as
 \begin{align}
		\frac{s}{N_c^2 T^3} =  \frac{\pi^2}{16} \left(3- y\right)^2 \left(1 + y\right).\nn
\end{align}
As stated by the holography principles, the entropy density of the boundary theory equals to the entropy density of the black brane. Likewise, the R-charge density of the boundary model $\rho=$ lim $_{\tilde{r}\to 0}\, \frac{\delta S}{\delta \Phi '}$ is given by
\begin{align}\label{rho}
		\frac{\rho}{N_c^2 T^3} =\frac{\sqrt{2}}{32} \sqrt{1-y^2}  (3 - y)^2.
	\end{align}
Having obtained the entropy and charge density with the aid of the thermodynamic equations $s=(\frac{\partial p}{ \partial T})_\mu$ and $\rho = (\frac{\partial p}{\partial \mu})_T$, one can derive the pressure of the dual strongly coupled theory 
\begin{equation}\label{p}
\frac{p}{N_c^2T^4}= \frac{\pi^2}{128} (3 - y)^3 (1+ y).
\end{equation}
The conformal symmetry of the underlying boundary theory leads to a relation between the energy density  $\varepsilon$ and pressure, namely $\varepsilon = 3p$ which is nothing but the tracelessness of the stress tensor.
On the other hand, the  Hessian matrix of the boundary thermodynamic quantities is 
\begin{align}
		\frac{\partial(s, \rho)}{\partial(T, \mu)} = \bigg(\begin{array}{cc}
		\frac{\partial s}{\partial T}& \frac{\partial s}{\partial \mu}\\
		\frac{\partial \rho}{\partial T}& \frac{\partial \rho}{\partial \mu}
		\end{array}\bigg),\nn
	\end{align} 
whose Jacobian $\mathcal{J}$, the determinant of the above matrix, is given by
\begin{align}
		\frac{\mathcal{J}}{N_c^4 T^4} = \frac{3\pi^2 }{256} (3-y)^4 (1 + \frac{1}{y}).\nn
	\end{align}
Clearly, the Jacobian is divergent at 	$y=0~(\frac{\mu}{T} = \frac{\pi}{\sqrt{2}})$  signaling that there is a second order phase transition. The  thermodynamics of the 1RCBH model is reviewed comprehensively  near the critical point, $y=0$, in the Ref. \cite{Finazzo:2016psx}. Moreover, it is easy to show that at $y=1$ we recover the thermodynamics of the $\mathcal{N}=4$ SYM at zero chemical potential.

\section{Hydrodynamics dual to the 1RCBH model} \label{sec:hydro-1RCBH}

We will extend the previous studies of the hydrodynamics dual to the 1RCBH model \cite{DeWolfe:2011ts} to the higher orders of gradient expansion. There are two distinct computational ways to study the hydrodynamics of the current model which, in principle, at the end of the day should lead to the same results. The first method,  is to use  the gauge fixing process, following KS \cite{Kovtun:2005ev}, and classify the perturbations according to the diffeomorphism and gauge transformations. On top of that, the little $SO(D-2)$ group can be used to specify various sectors of the perturbations.
The second method, is to adopt the master equations formalism \cite{Kodama:2003jz, Kodama:2003kk, Jansen:2019wag}.\footnote{We thank  Andrzej Rostworowski for constructive discussion on master equations formalism.} Unlike the KS method, in master formalism one should write the gauge invariant combination of all perturbations before using any gauge fixing which leads to gauge invariant linearized equations. There exist particular combinations of fields such that the equations of motion reduce to  the Schrodinger type equations. While for spin-2 sector there is no difference between those methods, the latter method has great advantages and  simplifies vastly the equations in other sectors.
In this paper, we study the hydrodynamics in  spin-2 and spin-1 sectors, utilizing the master formalism \cite{Kodama:2003jz, Kodama:2003kk, Jansen:2019wag}, and leave the spin-0 sector to the future works.

We close this part by addressing the importance of total action. As stated in  section \ref{sec:GreensF}, the transport coefficients are derived from two-point functions which are nothing but the derivatives of on-shell total action with respect to the sources. The total action is 
\begin{align}
	S_\textrm{tot} &= S_\textrm{bulk} + S_\textrm{GH} + S_\textrm{ct}, \nn
\end{align}
where $S_\textrm{bulk}$ is  the bulk action given in Equation \eqref{1R-action},  $S_\textrm{GH}$ is the Gibbons-Hawking boundary term and $S_\textrm{ct}$ is the counter term \cite{Critelli:2017euk},
\begin{align}\label{eq:GH-ct}
S_\textrm{GH}&=\frac{1}{8 \pi G_5}  \int d^4x \sqrt{-\gamma} K,\\
S_\textrm{ct} &= \frac{-1}{16 \pi G_5} \int_{\partial M} d^4 x \sqrt{-\gamma} \bigg[3 + \frac{R}{4}  + \frac{\log \tilde{r}}{16} \left(R^{\mu \nu} R_{\mu \nu} - \frac{R^2}{3}  + f(
\phi)  F_{\mu \nu} F^{\mu \nu}\right)\nn\\ 
&\hspace{125pt} + \frac{\phi^2}{2} \left( \frac{1}{\log \tilde{r}}-1\right)\bigg],\nn
\end{align}
where $\gamma_{ij}$ is the boundary induced metric, $K$ is the corresponding  extrinsic curvature, $K = \frac{1}{2} \gamma^{i j} \partial_r \gamma_{ij}$. We would like to emphasize that  the $S_\textrm{ct}$ contributes to the second order transport coefficients $\kappa$ and $\tau_\Pi$, while it has no impact on the first and third order ones.

\subsection{Spin-2 sector}\label{sec:spin-2}

We start by investigating the spin-2 sector which includes only one, namely perturbation $h_{xy}(t,z,r)=g_{xx}(r) {H}_{xy}(r) e^{-i \omega t+i q z}$ in which we use the SO(3) symmetry of the background to fix the momentum along $z$ direction. The linearized Einstein equation for this perturbation in Schwarzschild coordinate is given by 
\begin{align}\label{eq23}
	&H_{xy}''(r) - \left(\frac{1}{r} + 2 \frac{1+M^2 r^4}{r h(r) (1+Q^2 r^2)}\right)H_{xy}'(r) - 4 \frac{\mathfrak{q}^2 h(r) - \mathfrak{w}^2}{\alpha^2 h^2(r) (1+Q^2 r^2)} H_{xy}(r)=0,\nn\\
	& h(r)=1-\frac{M^2r^4}{1+Q^2r^2},\qquad \alpha\equiv\frac{2\sqrt{1+Q^2}}{2 + {Q^2}},
\end{align} 
where we  use the dimensionless frequency and momentum defined as
\be
\mathfrak{w}\equiv \frac{\omega}{2 \pi T},\qquad \mathfrak{q}\equiv\frac{q}{2 \pi T}.\nn
\ee 
The Equation \eqref{eq23} is a subtle linear second order differential equation which is very hard to find the exact analytical solutions. However, we are interested in the hydrodynamic limit, i.e. $\mathfrak{w}\ll1$ and $\mathfrak{q} \ll 1$. The solution representing the ingoing wave at the horizon $r=1$ can be written as (for more details see Appendix \ref{AppA}),  
\begin{eqnarray}
H_{xy}(r) = (1-r^2)^{-i \frac{\mathfrak{w}}{2}} \frac{\mathcal{H}(r)}{\mathcal{H}(0)},\nn
\end{eqnarray}
where $\mathcal{H}(r)$ is regular at the horizon.
In hydrodynamic limit we use the following ansatz
\begin{align}\label{eq30}
	\mathcal{H}(r)=\sum_{i=0}^{\infty}\epsilon^i\mathcal{H}_i(r),\qquad \qquad \left(\mathfrak{w}, \mathfrak{q}\right) \to \left(\epsilon \mathfrak{w}, \epsilon \mathfrak{q}\right),
\end{align}
to solve  Equation \eqref{eq23} order by order in $\epsilon$. The explicit form of the solutions up to third order is given in Equation \eqref{eqd3}. 

The next step is to compute the on-shell action and keep all the  terms quadratic in $H_{xy}$  following the recipe explained in Ref. \cite{Policastro:2002se} which leads to
\begin{align}\label{eq50}
 S_\textrm{tot} &=\frac{ \pi^2 N_c^2 T^4 \alpha^4}{8}\bigg[ \frac{(1-r^2)(1+M^2r^2)}{r^2} H_{xy}'(r) H_{xy}(r)\nn\\
 &\hspace{110pt}+ \left(\frac{\left(Q^2+2\right)^2(\mathfrak{q}^2 - \mathfrak{w}^2)}{4 M^2 r^2 }-\frac{M^2}{2}\right)H_{xy}^2(r)\bigg]_{r\to 0}.
\end{align}
We would like to highlight a practical point at this stage in our computation.
The contribution of the counter term in \eqref{eq50} starts from the second order which is given in the second line of Equation \eqref{eq50}. 
On the other hand, each term  in the counter term \eqref{eq:GH-ct} is composed of  even number of derivatives. Therefore, the general form of $S_\textrm{tot}$ is the same in $2k$ and $2k+1$ order of expansion.  
Since we are interested in computing the transport coefficients up to the third order, we only need to plug the solution of $H_{xy}$  to the corresponding order in \eqref{eq50}.
The retarded Green's function can be obtained by taking the second derivative of this total action with respect to the source \cite{Policastro:2002se},
\begin{align}
G^{R}_{xy, xy} &= \frac{N_c^2 \pi^2 T^4}{4} \bigg[ \frac{(3-y)^3 (1+y)}{32} - \frac{2 i \mathfrak{w} \alpha^4}{1+y} - \frac{4 \alpha^4 \mathfrak{q}^2}{(1+y) (3-y)}\nonumber\\
& + \frac{4 \mathfrak{w}^2 \alpha^4\left(1 - 2 \log2 + \log (1+y)\right)}{(3-y) (1+y)} - \frac{i \mathfrak{w}^3}{8} \lambda_{17}^{(3)} + i \mathfrak{w} \mathfrak{q}^2 (\lambda_1^{(3)} - \lambda_{16}^{(3)} - \lambda_{17}^{(3)})\bigg].\nn
\end{align}
Now we are in the position to apply the Kubo formula \eqref{eq37b} and compute the transport coefficients up to the third order of gradient expansion,
\begin{align}\label{eq446}
\eta & =  \frac{\pi N_c^2T^3 (3-y)^2(1+y)}{64},\nn\\
\kappa & =\frac{N_c^2 T^2 (3-y)(1+y)}{32},\nonumber\\
\tau_{\pi}& =\frac{2 - 2 \log(2) + \log (1+y)}{\pi T (3-y)},\\
\lambda_{17}^{(3)}&=\frac{N_c^2 T (1+y)}{128 \pi} \bigg(\log (\frac{4}{1+y}) \left(8 + (1+y) \log (\frac{1+y}{4})\right) + 2 (3-y) \text{Li}_2(\frac{3-y}{4})\bigg),\nonumber\\
\lambda_1^{(3)} - \lambda_{16}^{(3)} &= \frac{N_c^2 T (1+y)}{128 \pi} \bigg(\log (\frac{4}{1+y}) \left(16+ (1+y) \log (\frac{1+y}{4})\right) + 2 (3-y) \text{Li}_2(\frac{3-y}{4})\bigg).\nn
\end{align}

It is easy to check that the shear viscosity satisfies the universal relation  
\begin{align}\label{eq36}
	 \frac{\eta}{s} = \frac{1}{4 \pi},\nn
	\end{align}
which is in complete agreement with \cite{Policastro:2001yc,DeWolfe:2011ts}.
Moreover the $y=1$ limit of those transport coefficients coincide with the counterparts of the $\mathcal{N}=4$ SYM theory dual to AdS$_5$ black brane background \cite{Grozdanov:2015kqa, Baier:2007ix}. 
One of the remarkable features of these transport coefficients  is their behaviour near the critical point of the phase transition.
In the section \ref{sec:critical} we will show that all the transport coefficients  reach to their critical value with the same critical exponent $\theta = \frac{1}{2}$. 

\subsection{Spin-1 sector}\label{sec:spin-1}

In this section we  study  the hydrodynamic limit corresponding to the spin-1 sector. In our setup, by employing the master equations formalism \cite{Jansen:2019wag} the coupled equations of motion  of the gauge invariant perturbations can be written in a  decoupled  form. Indeed the chief advantage of using this approach is twofold. First, it authorizes the analytical investigations of this sector, as we will discuss in this section. Second, the numerical computation of the QNM frequencies will be less costly. We will come back to this point in section \ref{sec:QNM}.

Let us summarize the key steps of the master equations formalism \cite{Kodama:2003jz,  Jansen:2019wag, Kodama:2003kk} in the following. At the first step, we should find the gauge-invariant combinations of the perturbation in spin-1 sector with plane wave ansatz $\Phi(r)e^{-i\omega t+i q z}$  for all the fields
\bea
\mathcal{Z}_1(r)\equiv h_{tx}(r)+i \omega h_{zx}(r),\quad \mathcal{Z}_2(r)\equiv h_{rx}(r)- h_{xz}'(r)+2A'(r)h_{xz}(r),\quad \mathcal{Z}_3(r)\equiv a_x(r).\nn
\eea 
The next step is to write the linearized equations of motion for the gauge invariant perturbations $\mathcal{Z}_i~(i=1, 2, 3)$ which are coupled equations. At the final step we should rewrite the gauge invariant combinations as linear combinations of the master scalars and their derivatives such that they satisfy master equations. The master scalars corresponding to the spin-1 sector are defined by
\bea
&&\mathcal{Z}_1(r)\equiv \frac{r^{2}  h(r)  e^{3 A(r) -B(r) }}{q^2} \left(3A'(r) \Psi_2(r)   +\Psi_2'(r) \right),\nn\\
&&\mathcal{Z}_2(r)\equiv-i\omega \frac{ e^{A(r) +B(r) }}{q^2 r^{2} h(r) }\Psi _2(r),\nn\\
&&\mathcal{Z}_3(r)\equiv \frac{e^{A(r) } }{q \sqrt{f(\phi(r)  )}} \Psi _1(r),\nn
\eea
and the master equations are given by 
\be\label{eq:master-eqs}
\square \bigg(\begin{array}{c}
	 \widetilde{\Psi}_1\\
	 \widetilde{\Psi}_2
	 \end{array}\bigg) - 
	 \bigg(\begin{array}{cc}
	W_{1,1} & W_{1,2}\\
	W_{1,2} & W_{2,2}
	\end{array}\bigg) \bigg(\begin{array}{c}
	 \widetilde{\Psi}_1\\
	 \widetilde{\Psi}_2
	 \end{array}\bigg) =0,\nn
\ee
where $\widetilde{\Psi}_i\equiv\Psi_i(r)e^{-i \omega t + i q z}$, the d'Alembert operator $\square$ is the wave operator on the background and the potential matrix $W$ is given in terms of the background functions. 

While in general the final equations are still coupled they have a simple form in terms of the master scalars. Nevertheless, in our case, since the following condition holds 
\be
\frac{W_{1,1}-W_{2,2}}{W_{1,2}}=-\frac{2 \alpha \sqrt{Q^2+1} }{q Q}=\text{constant}.\nn
\ee
The master equations \eqref{eq:master-eqs} can be further simplified to decoupled equations as
\bea\label{eq:master-eq}
&&\square \Psi_\pm \pm W_\pm \Psi_\pm=0,\\
&&W_\pm=\frac{r^2 \left(4 Q^4 r^2\left(M^2 r^4+1\right)+Q^2 \left(6 M^2 r^4+7\right)+5 M^2 r^2\right)+3}{\left(Q^2 r^2+1\right)^{7/3}}\pm\frac{4 M r^4 \sqrt{M^2+\frac{\mathfrak{q}^2 Q^2}{\alpha ^2}}}{\left(Q^2 r^2+1\right)^{4/3}}.\nn
\eea
This form of master equations make our analytical studies more intelligible. It turns out that the equation with $"+(-)"$ sign describes the shear channel perturbation  (transverse perturbation of the gauge field) . In the rest of this section we solve Equation \eqref{eq:master-eq} with $"+"$ sign perturbatively to find the hydrodynamic dispersion relation of the shear mode following the prescription given in Ref. \cite{Policastro:2002se}. We solve the relevant equation in Eddington-Finklestein (EF) coordinate by using the following ansatz 
\bea
\Psi_+(r)=r^{1/2}\sum_{n=0}\epsilon^n\psi_n(r),\nn
\eea
and we scale the frequency and momentum as $(\mathfrak{w}, \mathfrak{q}) \to (\epsilon^2 \mathfrak{w}, \epsilon \mathfrak{q})$. Again we utilize the variation of parameters method reviewed in Appendix \ref{AppA} to find the $\psi_n$'s. At the zeroth order in $\epsilon$ one can solve a second order differential equation for $\psi_0$ with regularity condition at the horizon
\be
\psi_0(r)=\frac{M\ r}{\sqrt{Q^2 r+1}},\nn
\ee
where the second integration constant is fixed by imposing  $\psi_0(1)=1$. In higher orders, without loose of generality, we impose $\psi_i(1)=0$ for $i>0$. We  find the solutions analytically up to the second order and the source-less boundary condition leads to the following form of the spectral curve \citep{Grozdanov:2019uhi} in shear channel 
\bea
&&F_\textrm{shear}(\mathfrak{\mathfrak{q}}^2,\mathfrak{w})=
 \mathfrak{w}+\frac{i \mathfrak{q}^2 \left(Q^2+2\right)}{4 \left(Q^2+1\right)}-\frac{i \mathfrak{q}^4 \left(Q^2+2\right)^4}{64 \left(Q^2+1\right)^3}\nn\\
&&-\frac{i \mathfrak{w} ^2 \left[Q^2 \log \left(4 Q^2+4\right)+2 \log \left(2 Q^2+4\right)+2 \sqrt{Q^2+1} \cot ^{-1}(\sqrt{Q^2+1})-2 \coth ^{-1}\left(2 Q^2+3\right)\right]}{2 \left(Q^2+1\right)}\nn\\
 &&+\frac{\mathfrak{w}  \mathfrak{q}^2  \left(Q^2+2\right) \left[\sqrt{Q^2+1} \log \left(4\frac{Q^2+1}{Q^2+2}\right)+2 \cot ^{-1}\left(\sqrt{Q^2+1}\right)\right]}{8 \left(Q^2+1\right)^{3/2}}+\mathcal{O}(\mathfrak{w}^3, \mathfrak{w}^2 \mathfrak{q}^2, \mathfrak{w} \mathfrak{q}^4, \mathfrak{q}^6)=0.\nn
\eea

By solving this equation perturbatively in $\mathfrak{q}$, one can find the dispersion relation of the shear hydrodynamic mode 
\be\label{eq:hydro-shear}
\mathfrak{w}=-i\frac{ 1}{3-y}\mathfrak{q}^2-i \frac{4 (y+1) \log (2)-2 (y+1) \log (y+1)-4}{(y-3)^3 (y+1)}\mathfrak{q}^4+\mathcal{O}(\mathfrak{q}^6),
\ee 
where we use  ${Q=\sqrt{\frac{2(1-y)}{1+y}}}$. By comparing our results with general form of the dispersion relation in the shear channel for conformal theories (see for example \cite{Grozdanov:2015kqa}),
\be
\omega=-i\frac{\eta}{\varepsilon+p}q^2+\left(\frac{\eta^2 \tau_\pi}{(\varepsilon+p)^2}-\frac{1}{2}\frac{\theta_1}{\varepsilon+p} \right)q^4+\mathcal{O}\left(q^6\right),\nn
\ee
one can find a relevant third order transport coefficient $\theta_1$ as
\be\label{eq:theta1}
\theta_1\equiv-(\lambda_1^{(3)}+\lambda_2^{(3)}+\lambda_4^{(3)})=\frac{N_c^2 T}{32\pi} y.
\ee
Note that at the critical point of the phase transition the $\theta_1$ vanishes. This could be a hint to the symmetry enhancement of the underlying theory at the critical point. One may compare this phenomenon with vanishing bulk viscosity for theories with conformal symmetry. We will elaborate on this point in the discussion section.
Let us close this section by emphasizing that without using the master equations formalism \citep{Jansen:2019wag} the above computation could not be achieved in this straightforward manner (as pointed out in a simpler case such as AdS-RN black holes \cite{Abbasi:2020ykq}). In principle, one can continue and find the higher order terms in the dispersion relation by solving ordinary second order differential equations for higher $\psi_n$ using the variation of parameters method.  

\section{QNMs and convergence of the hydrodynamic series }\label{sec:QNM}

In this section we compute the QNM frequencies in spin-2 and spin-1 sectors associated with the poles of the corresponding retarded Green's function of the boundary theory.
We consider the complex momentum square  $\mathfrak{q}^2=|\mathfrak{q}^2| e^{i \varphi}$ and compute the QNM frequencies. That is simply because of the symmetries of the static background which leads to the fact that the linearized equations are functions of even powers of momentum.

To compute the QNMs in our background we used the pseudo spectral Chebyshev discretization along the radial coordinate in  EF parameterization. The boundary conditions we should impose are ingoing wave at the horizon which translates to the regularity condition in EF coordinate, and sourceless Dirichlet boundary condition at the asymptotic region which comes from the holographic dictionary. In all  the cases we consider in this work, the 30 number of grid points along the radial coordinate are enough to find $\sim\# 10$ lowest eigenmodes and we compare the results with 60 number of grid points to select the reliable modes with $10^{-6}$ relative accuracy. To cross-cheek our numerical findings we compute the dispersion relation of the hydro mode in small real momenta in shear channel and compare it with the analytical result presented in Equation \eqref{eq:hydro-shear} which shows perfect agreement. 

\begin{figure}
    \centering
    \includegraphics[width=0.47\textwidth]{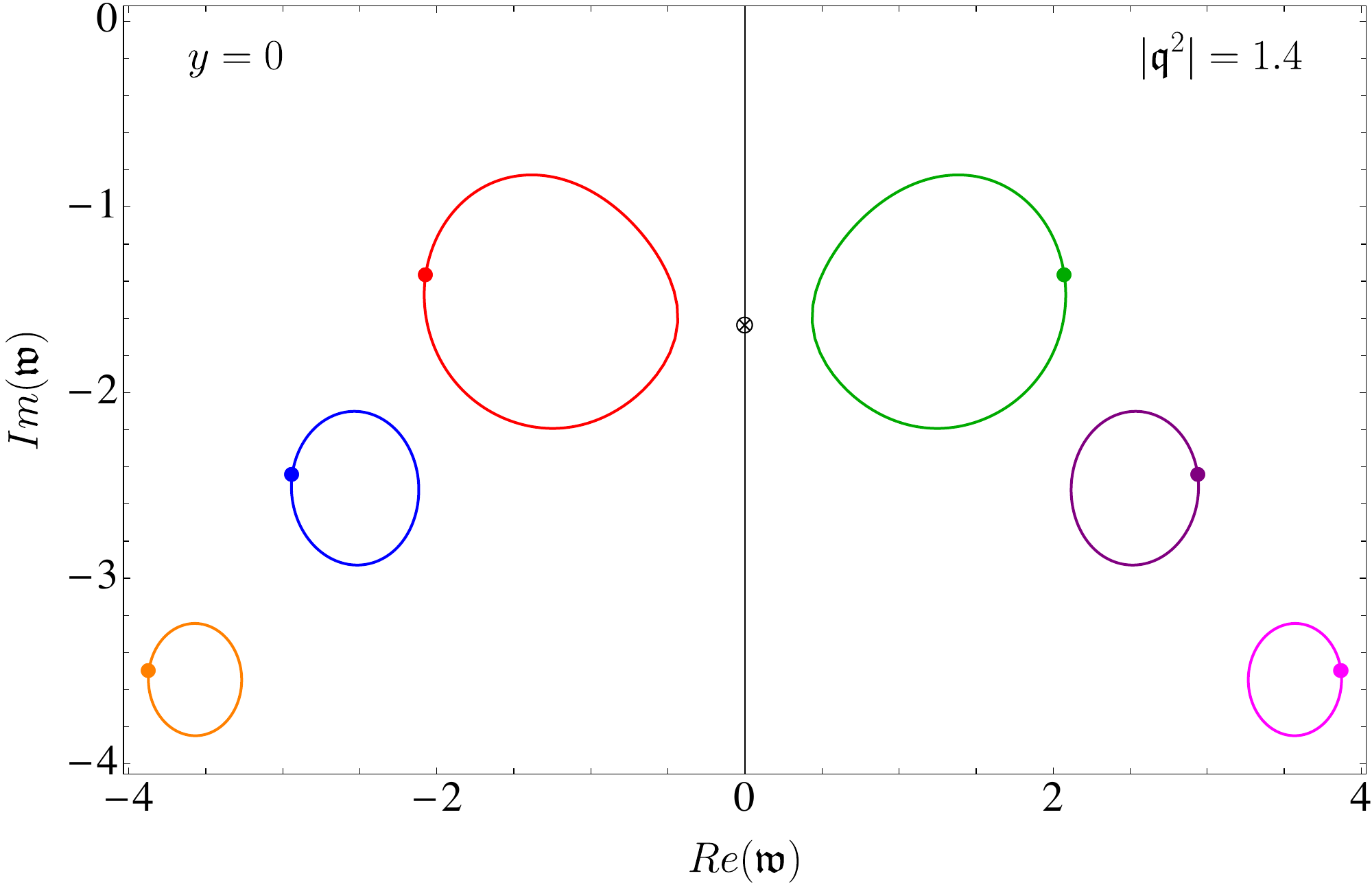}
    \includegraphics[width=0.47\textwidth]{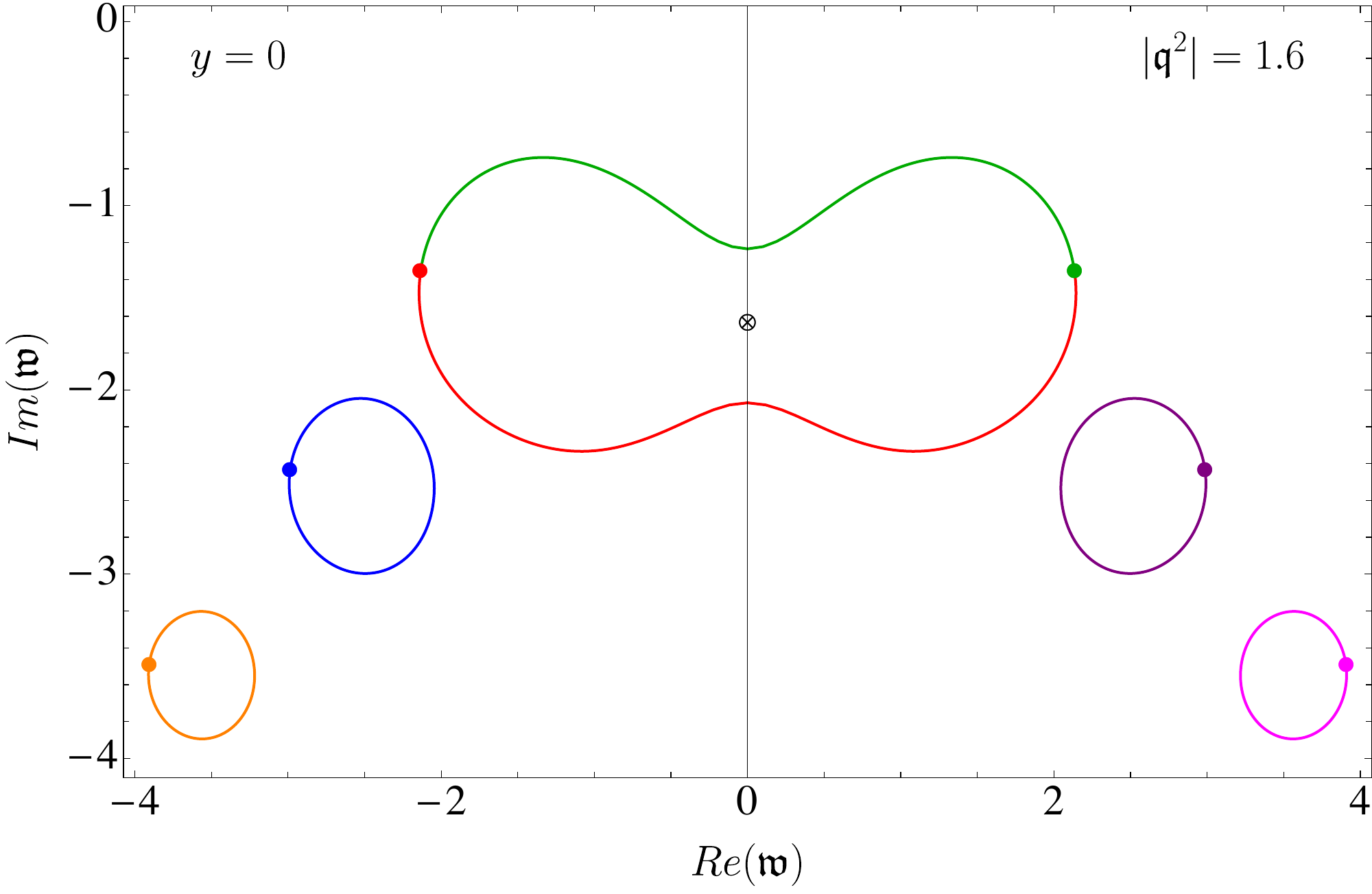}
    \includegraphics[width=0.47 \textwidth]{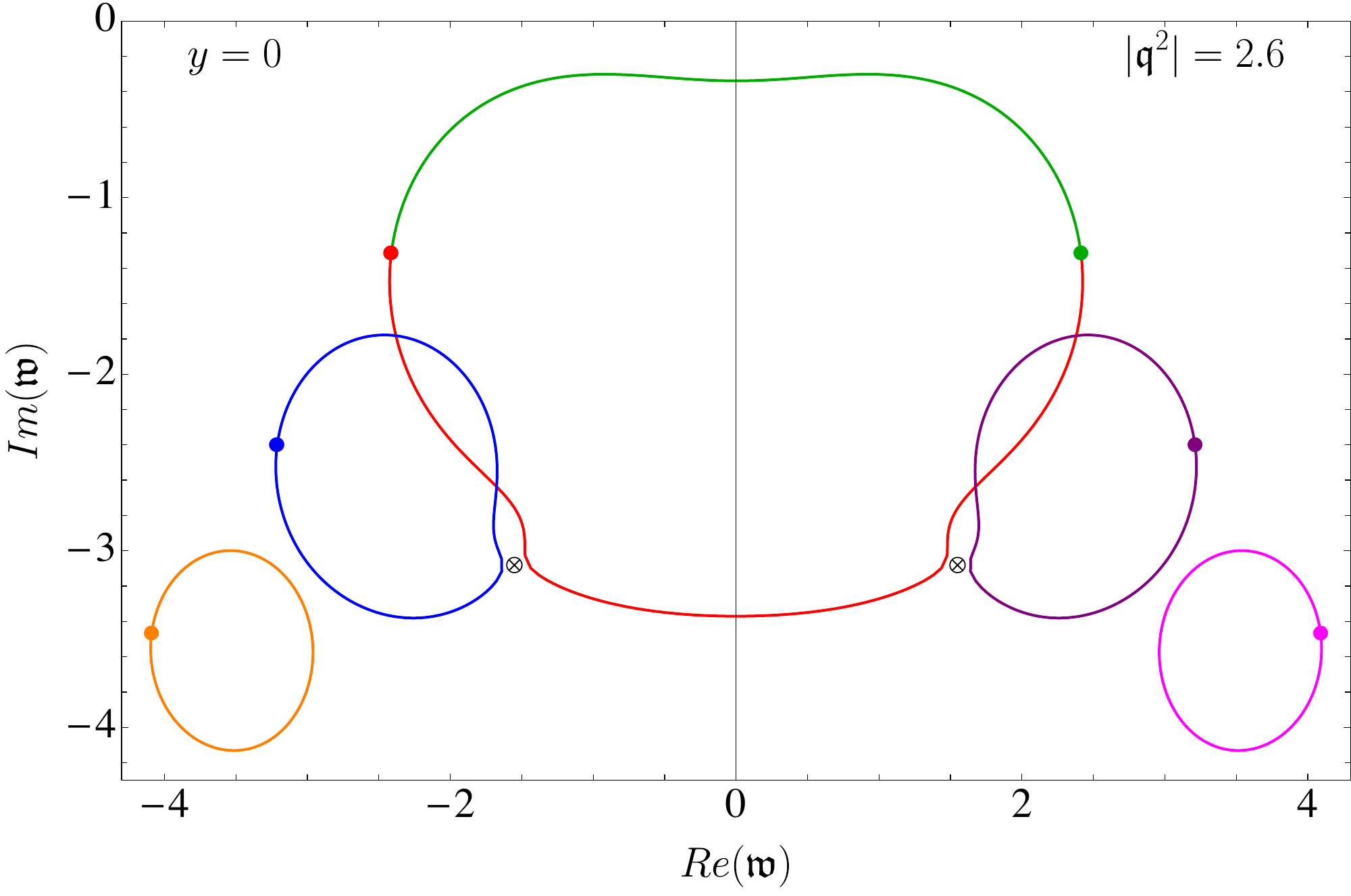} 
    \includegraphics[width=0.47 \textwidth]{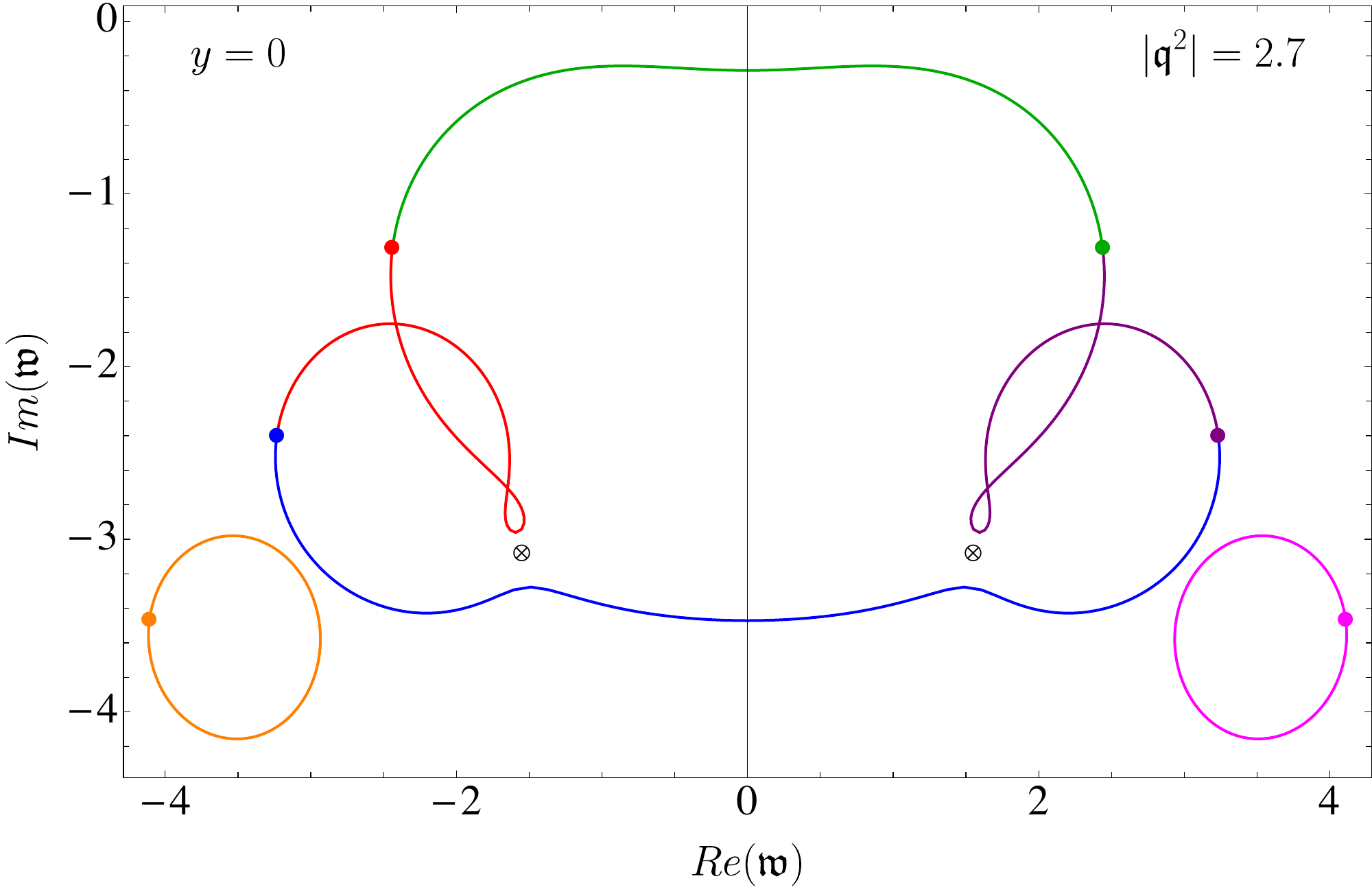}
    \caption{ The trajectory of the lowest QNMs in spin-2 sector  at the critical point ($y=0$). In the first row we show the trajectories before and after the first collision between the lowest modes at $|\mathfrak{q}_*^2|=1.505$ and $\mathfrak{w}_*=- 1.637 i$. 
    In the second row we show the trajectories before and after the second collision  at $|\mathfrak{q}_*^2|=2.621$ and $\mathfrak{w}_*=\pm 1.551 -3.081 i$.}
    \label{fig:qnm-scalar}
\end{figure}

Finding the radius of convergence  for series is an interesting topic in complex analysis \cite{ctcwall:67327ti}. 
Suppose we have  an analytic (spectral) curve $\mathcal{T}(u, v)=0$ of complex variables $(u, v)$ in the complex plane and we want to find $v_\star = v(u)$. These solutions are classified as regular points and critical points. Regular points are the zeros of curve which $\frac{\partial^i \mathcal{T}}{\partial v^i}|_{v = v_\star(u)} \neq 0$ for $i\geq 1$, while critical points of order $j$ are the zeroes which $\frac{\partial^i \mathcal{T}}{\partial v^i}|_{v = v_\star(u)} = 0$ for $1 \leq i \leq  j$. By definition, the radius of convergence for $v_\star$
is determined by the location of the nearest critical point to the origin. Therefore, at the critical point there is a degeneracy of solutions which specifies the radius of convergence. As explained in details in \cite{Heller:2020hnq}, if the spectral curve is non-analytic at some points then there may be other sources of singularities. 
In  all the cases  studied in this paper we did not find any sign of non-analyticity in the spectral curves for momenta smaller or equal to the critical momentum.\footnote{We thank Michal Heller for  pointing this out. }

\subsection{Spin-2 sector}\label{sec:QNM-s2}

In this section we consider the perturbations in the spin-2 sector and compute the corresponding QNMs numerically. Since there is no hydro-mode in this sector we can only find the radius of convergence of the non-hydro modes and we will focus only on the lowest modes.  
In small values of $|\mathfrak{q}^2|$ each mode has a closed trajectory for $\varphi$ from $0$ to $2\pi$ while for larger values they may collide and share their trajectories. Due to the symmetries the collision between the pair modes is always on the imaginary axes in the complex frequency plane with a purely imaginary momentum. The trajectories of the QNMs are qualitatively the same for the whole range of $y$. In Figure \ref{fig:qnm-scalar},  as an example, we illustrate the trajectories of the modes before and after the first collision for $y=0$. 

\begin{figure}
    \centering
        \includegraphics[width=0.47 \textwidth]{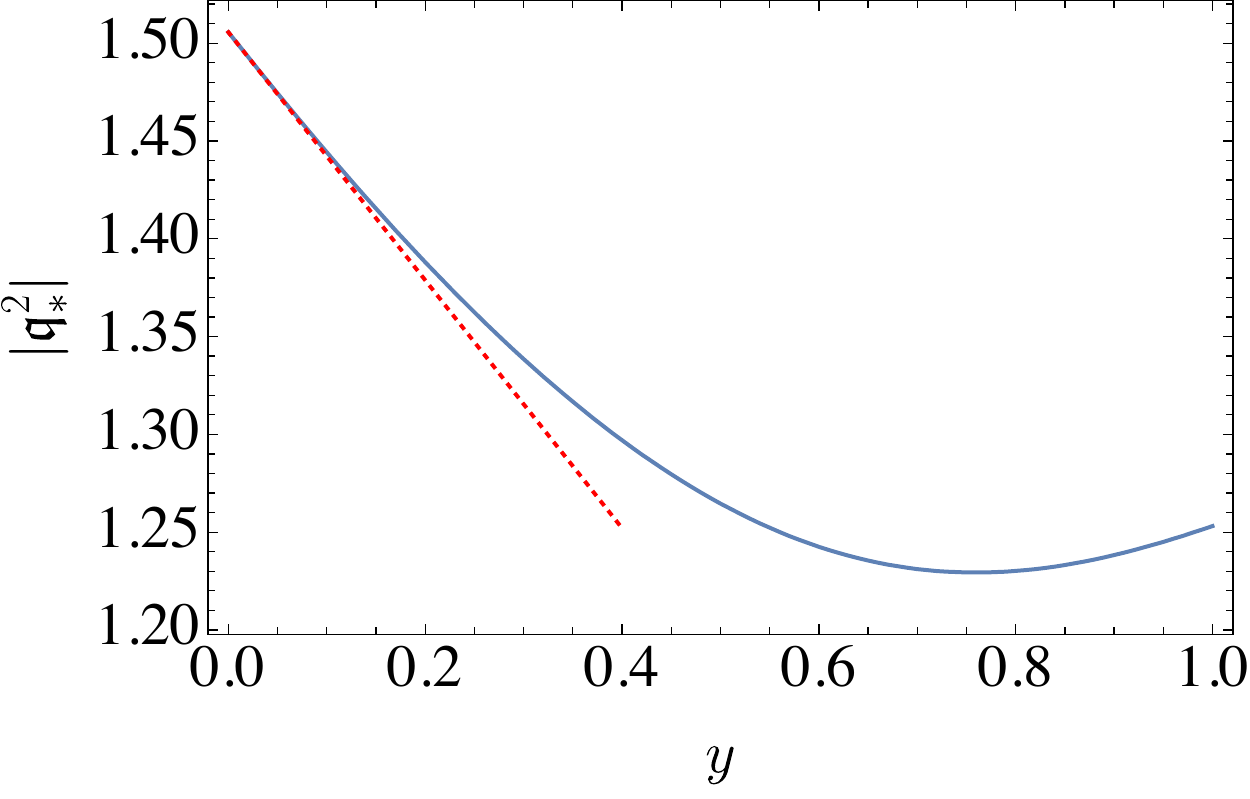} 
        \includegraphics[width=0.49 \textwidth]{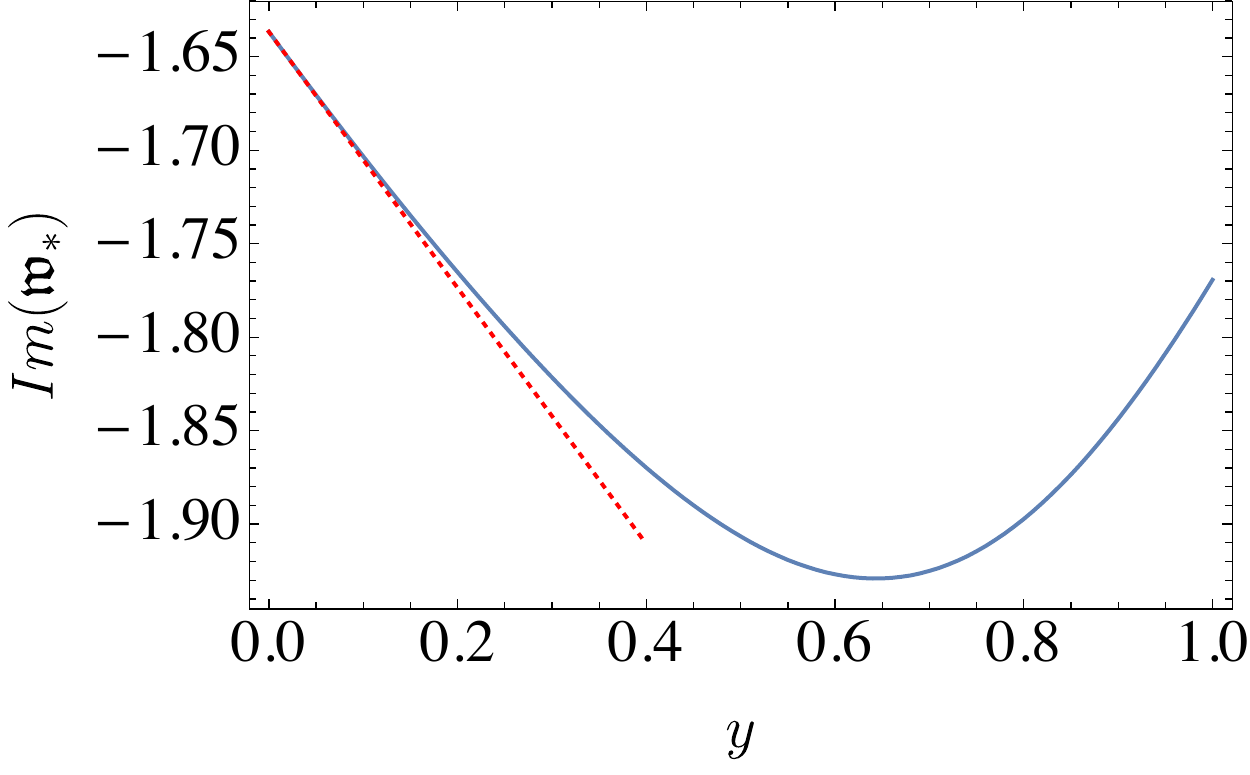} 
    \caption{Left panel: the $y$ dependency of the convergence radius of the lowest non-hydro modes in spin-2 sector computed numerically. Right panel: the corresponding frequency at which the lowest non-hydro modes collide. In both panels the solid blue lines are the results and the red-dotted lines are the fits close to the critical point given in Equation \eqref{eq:qc-spin-2}. }
    \label{fig:qnm-scalar-y-k}
\end{figure}

The general features of the lowest collision in this sector does not change in the whole range of $y$. In Figure \ref{fig:qnm-scalar-y-k} we show the radius of convergence of the lowest mode in the spin-2 sector and also the corresponding frequency as functions of $y$. Close to the critical point the radius of convergence and the corresponding frequency are linear functions in $y$ which can be fitted by
\be\label{eq:qc-spin-2}
|\mathfrak{q}_*^2| \simeq 1.505-0.634\, y, \qquad i\mathfrak{w}_* \simeq   1.637 +0.685\, y.
\ee
In section \ref{sec:critical} we will discuss on these relation.

\subsection{Spin-1 sector}\label{sec:QNM-s1}

In this section we consider the perturbations in the spin-1 sector, including the shear and the transverse gauge field channels, and compute the lowest QNMs using numerical techniques. 

In small values of $|\mathfrak{q}^2|$ each mode has a closed trajectory for $\varphi$ from $0$ to $2\pi$. In other words, in this regime each mode can be found uniquely in complex momentum square plane. On the other hand for larger values of $|\mathfrak{q}^2|$ this may change due to level-crossing or collision of the modes \cite{Grozdanov:2019kge}. In fact, depends on how far the medium is from the critical point either of phenomena may happen. In the rest of this section we present various plots of the trajectory of the QNM frequencies for the complex momentum square and discuss the main features at different regimes. One of the  purposes is to investigate the convergence radius of the  hydrodynamic series in the whole regime of our model. In particular its behaviour near the second order phase transition  will be explored in the next section.

\begin{figure}
    \centering
        \includegraphics[width=0.47 \textwidth]{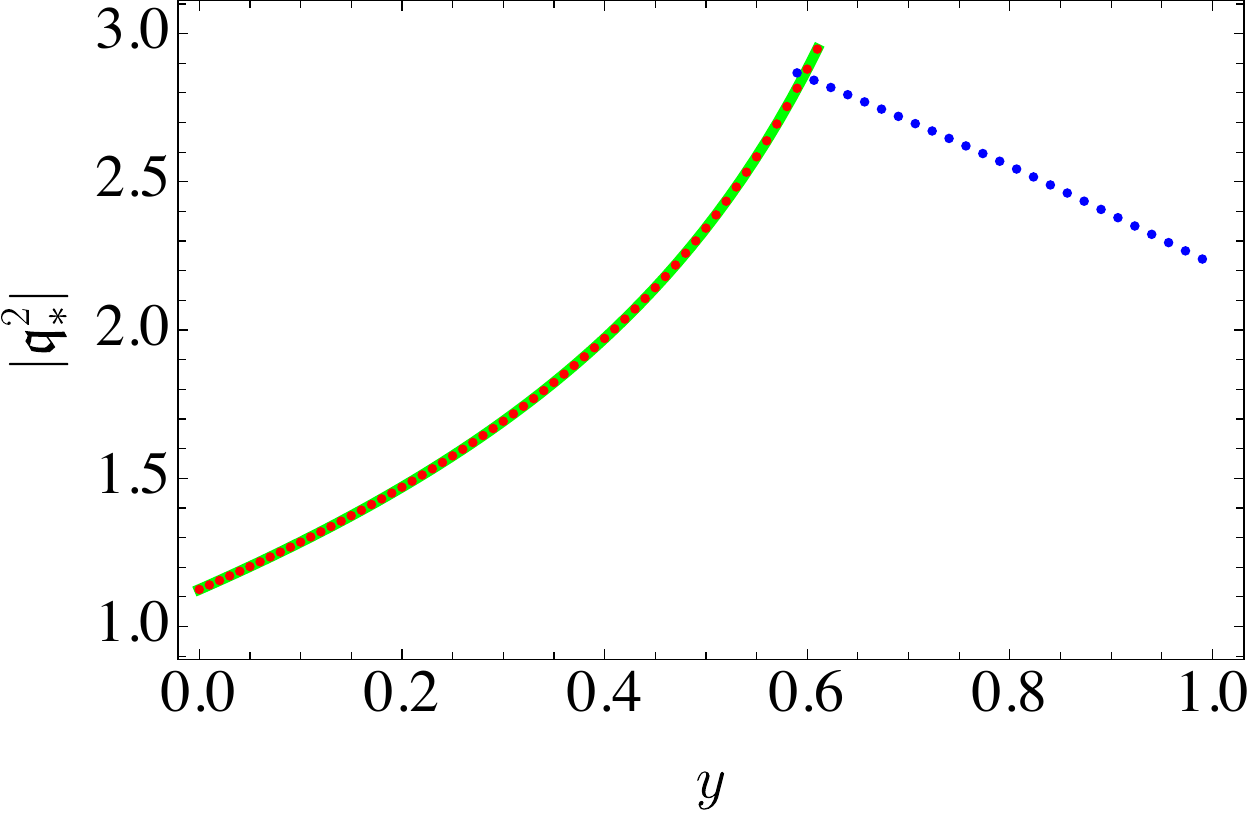} 
        \includegraphics[width=0.47 \textwidth]{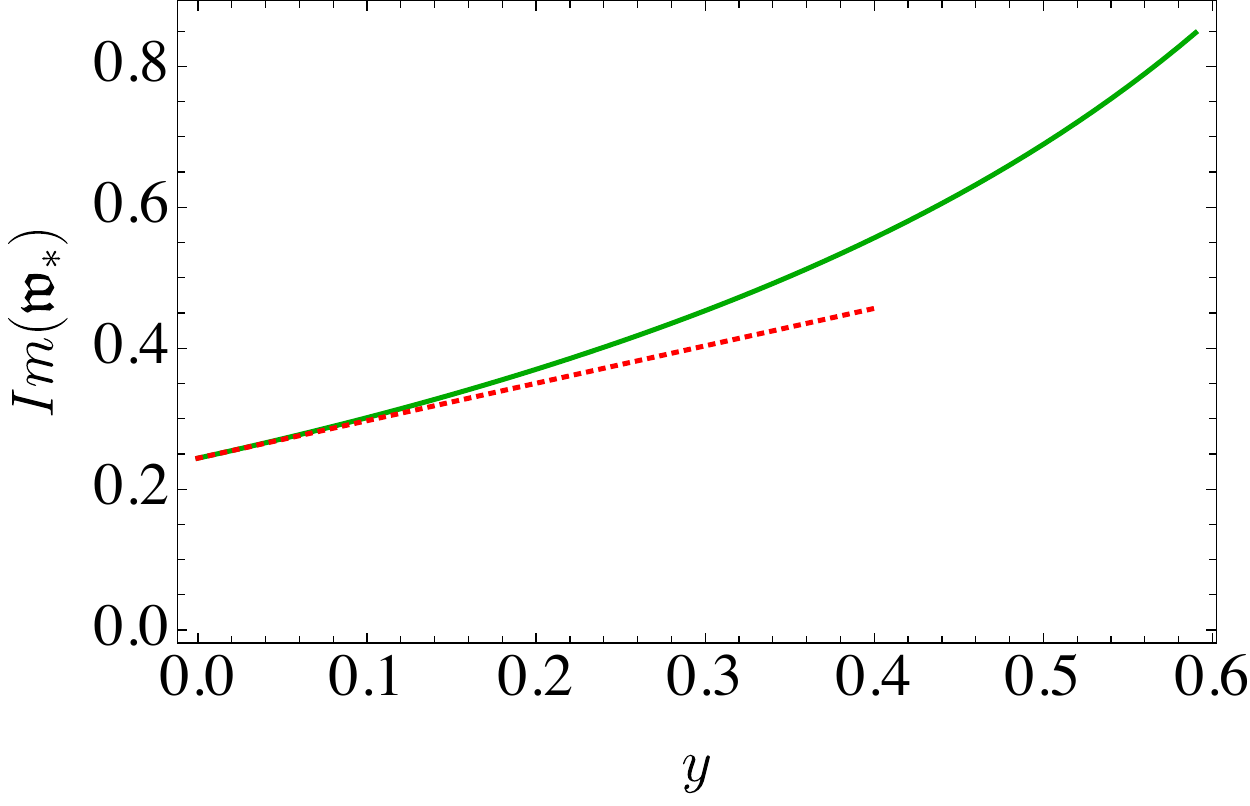} 
    \caption{Left panel: the y dependency of the convergence radius of shear mode. The red and blue dotted lines are our numerical computation and the green-solid line is analytical results given in Equation \eqref{qc-analytic}. Right panel: The corresponding frequency for $0\leq y \leq 0.596$ (solid green line) and the linear fit near critical point given in Equation \eqref{eq:qc-spin-1}.}
    \label{fig:qnm-shear-y-k}
\end{figure}

To compute the  QNMs of the system in complex momentum square, we employ  the master equations formalism \cite{Jansen:2019wag} explained in  section \ref{sec:spin-1}.  The decoupled linearized equations are given in Equation \eqref{eq:master-eq} and as pointed out in \cite{Withers:2018srf,Jansen:2020hfd} there is a collision of the modes due to the appearance of a radical in the last term of the potentials $W_\pm$. If the lowest collision of the hydrodynamic mode  occurs at this $|\mathfrak{q}^2|$, then the radius of convergence can be related to this phenomena,
\be\label{qc-analytic}
|\mathfrak{q}_*|^2=-\frac{\alpha^2M^2}{Q^2}=\frac{(3-y)^2 (y+1)}{8 (y-1)}.
\ee
It is easy to show that the critical momentum given in Equation \eqref{qc-analytic} has a simple expression in terms of the thermodynamic variables
  \begin{align}
     q_*=i\frac{\varepsilon+p}{\pi \rho}.
  \end{align}
This branch point is of square-root type and at this value of the momentum the shear and transverse gauge modes  satisfy the same equations of motion. Therefore, for each mode in the former there is a cousin in the latter which they meet at the critical momentum. In other words, the convergence radius of the hydrodynamics is constrained by a crossing between the hydrodynamic mode and one of the modes in the  transverse gauge channel.

\begin{figure}
    \centering
    \includegraphics[width=0.47\textwidth]{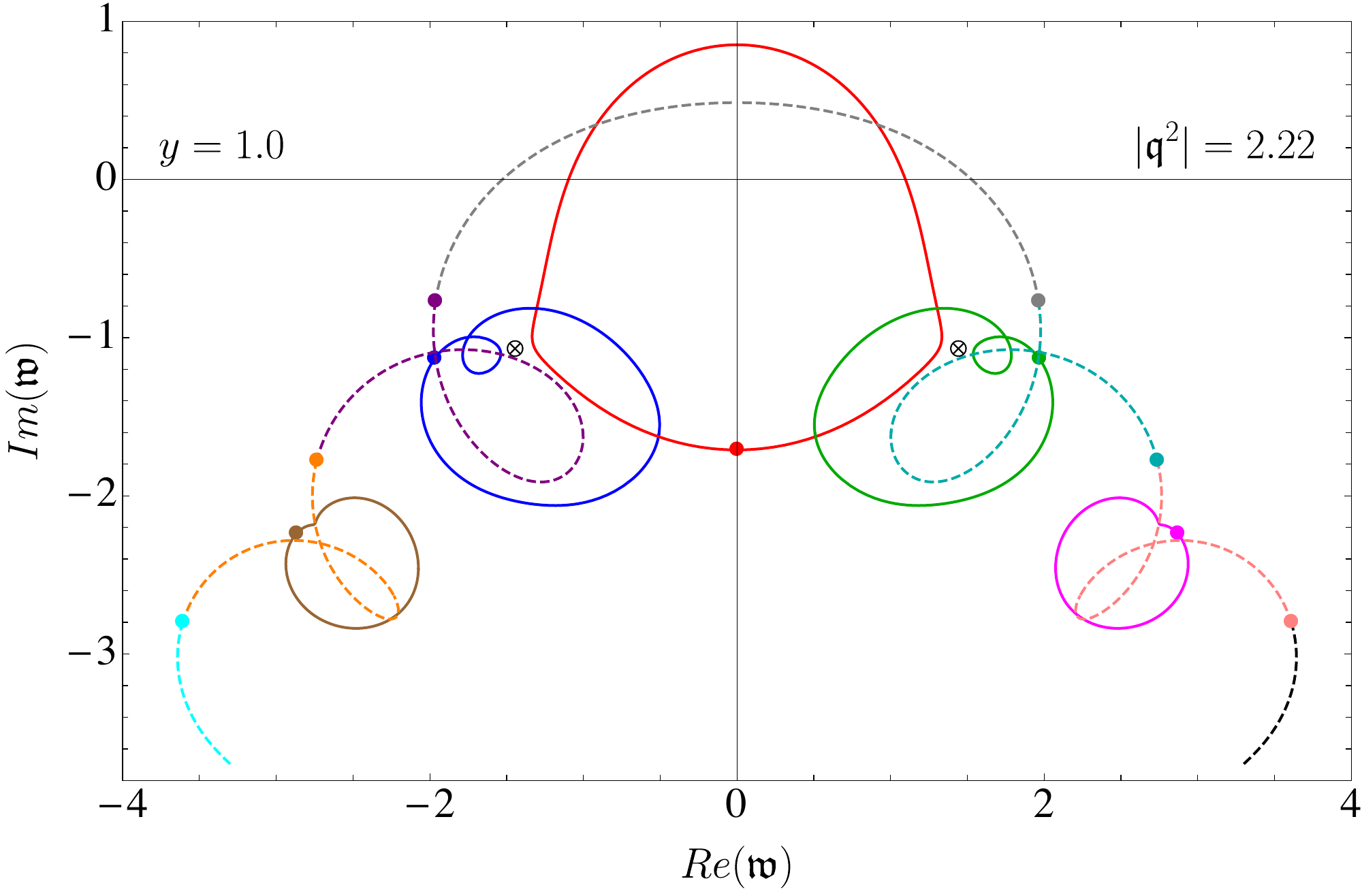}
    \includegraphics[width=0.47\textwidth]{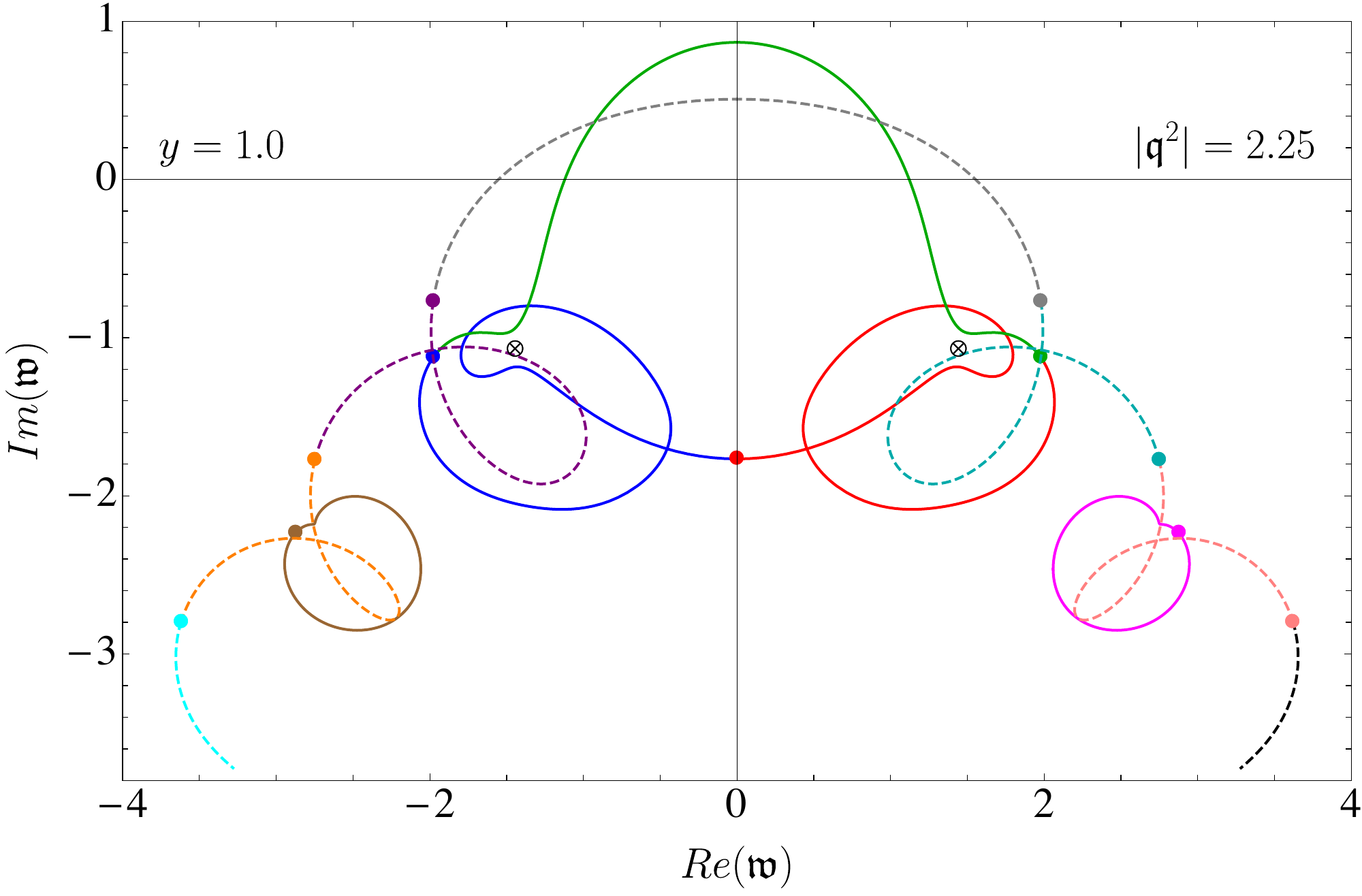}
    \includegraphics[width=0.47 \textwidth]{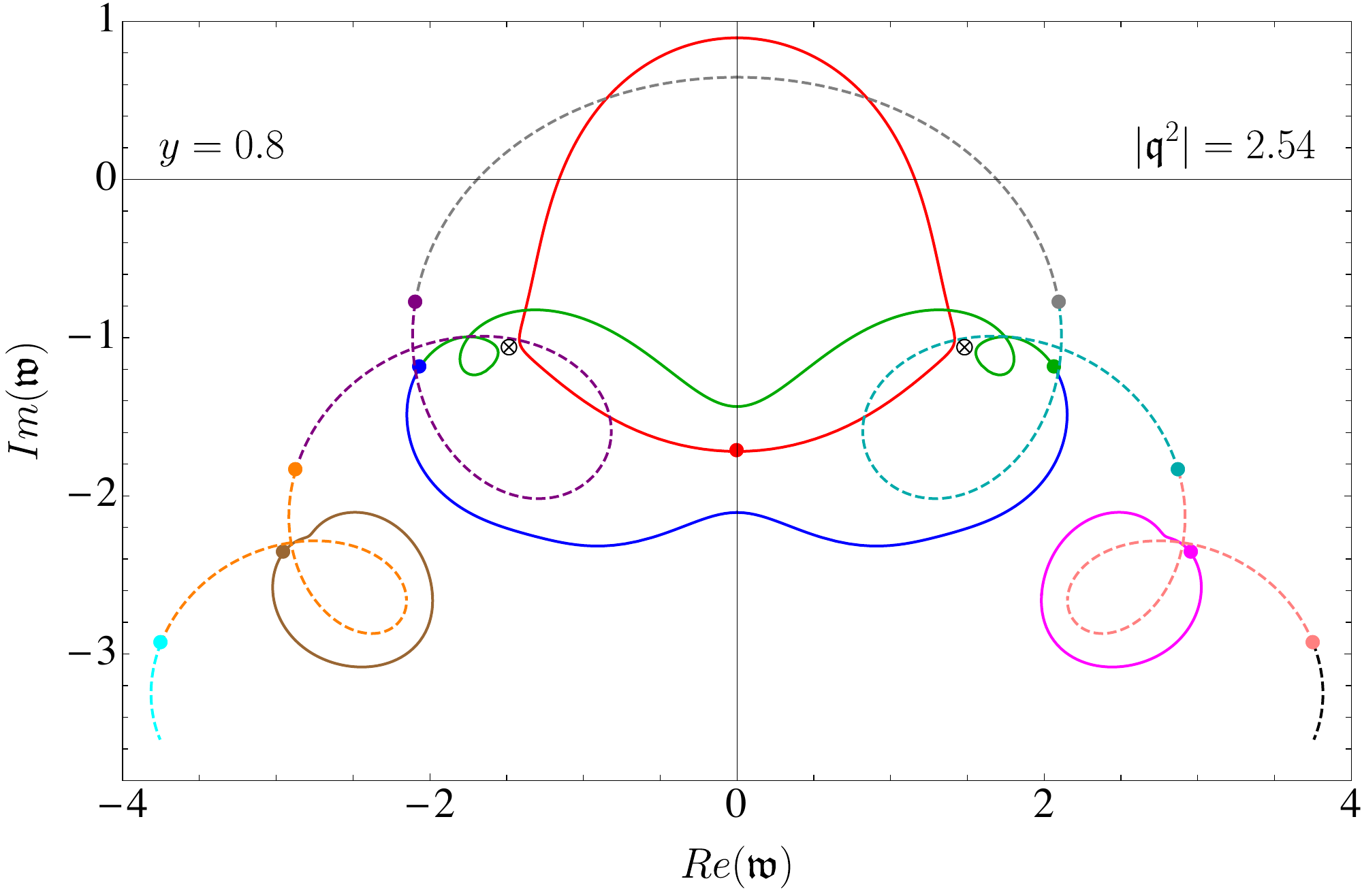} 
    \includegraphics[width=0.47 \textwidth]{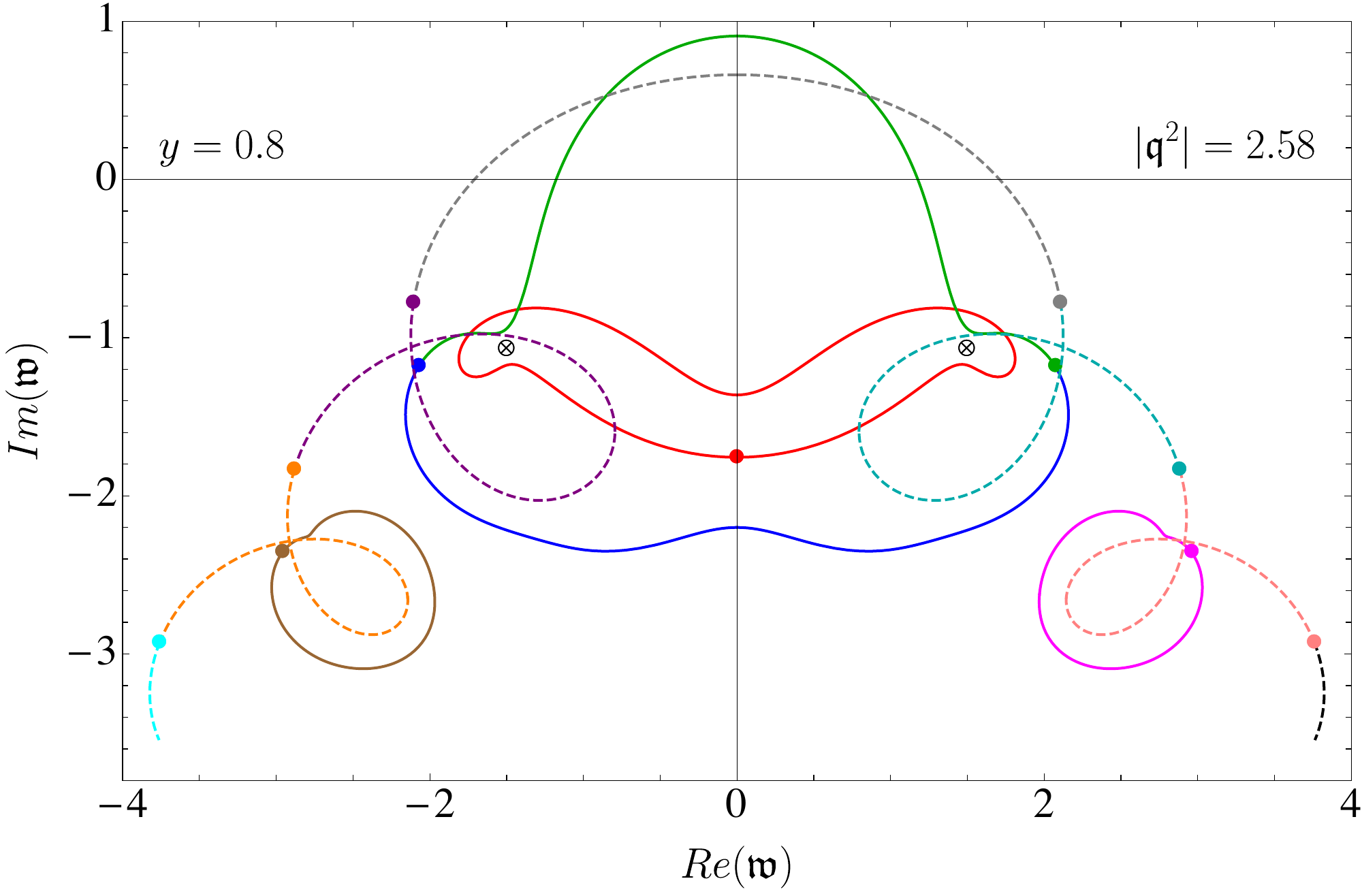}
    \caption{The trajectory of the lowest QNMs in shear channel (solid lines) and in the transverse channel of the gauge field (dashed lines)  for $y=1$ (AdS-Schwarzschild black brane) in the first row and for $y=0.8$ in the second rows. Top-left (right) panel: the trajectory of the lowest modes before (after) the joining of the hydro mode to the lowest non-hydro shear mode.  The joining happens at $|\mathfrak{q}_*^2|=2.224$. Bottom-left (right) panel: the trajectory modes before (after) the collision  occurs at $|\mathfrak{q}_*^2|=2.553$. The collision of hydro and  the lowest non-hydro shear mode happen leads to exchange of their trajectories and not joining to a common trajectory. This is so called level-crossing \cite{Grozdanov:2019uhi}.}
    \label{fig:qnm-shear-y-1}
\end{figure}

Our main results can be summarized in Figure  \ref{fig:qnm-shear-y-k}  in which  the $y$ dependency of the convergence radius is demonstrated, where it indicates that there exist at least two different structures for the first mode collision of the hydrodynamic QNM. In the left panel, the red and blue points are what we found by studying the QNMs of the system in $\mathfrak{q}^2$ numerically while the solid green line is the analytical formula for the convergence radius \eqref{qc-analytic} computed by employing  the master equations formalism \cite{Jansen:2019wag}. In the right panel we show the corresponding frequency for $0\leq y \leq 0.596$ (solid green line) and the linear fit near critical point given by
\be\label{eq:qc-spin-1}
-i\mathfrak{w} = 0.244 + 0.531\, y.
\ee

\begin{figure}[t]
    \centering
    \includegraphics[width=0.47 \textwidth]{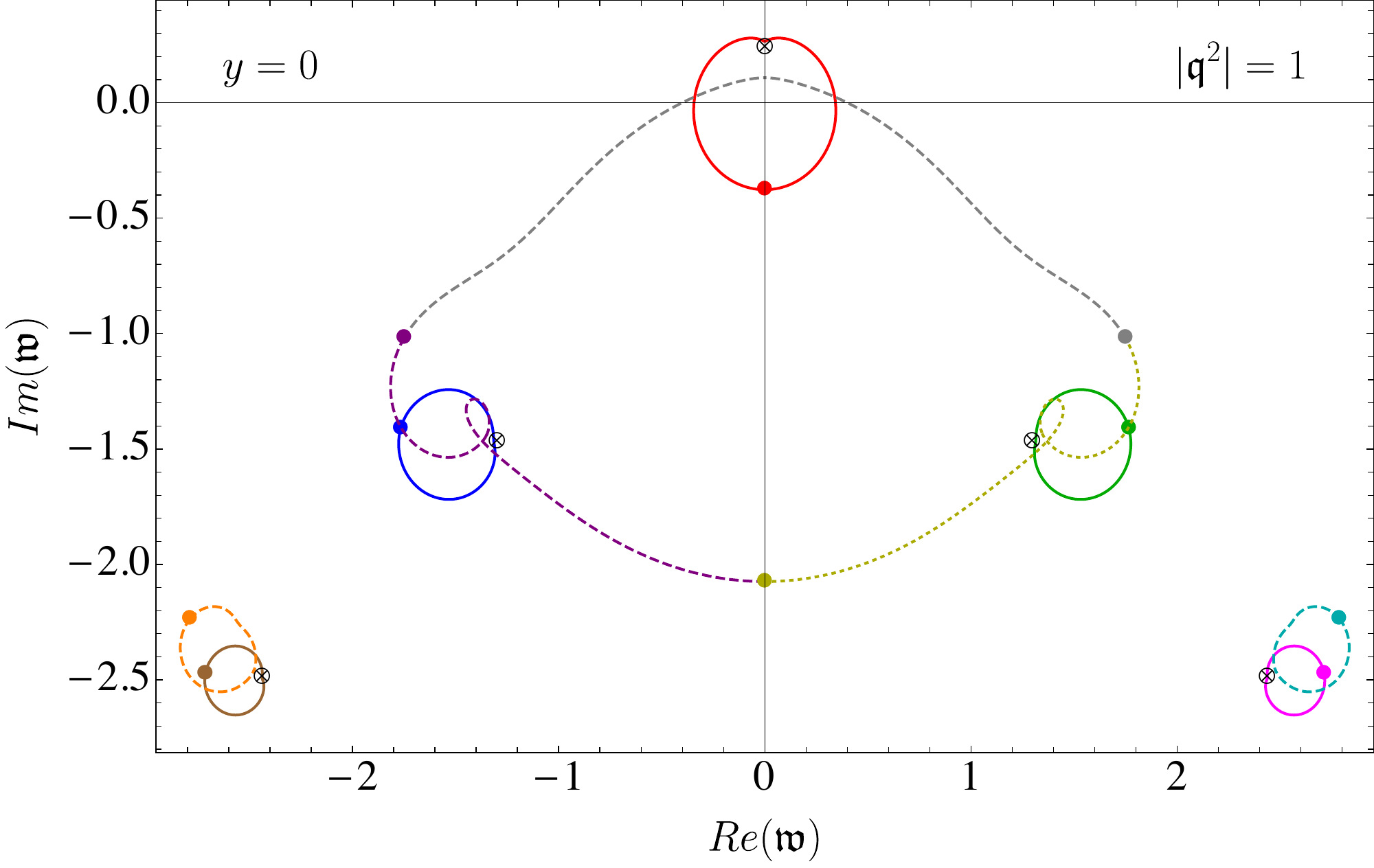} 
    \includegraphics[width=0.47 \textwidth]{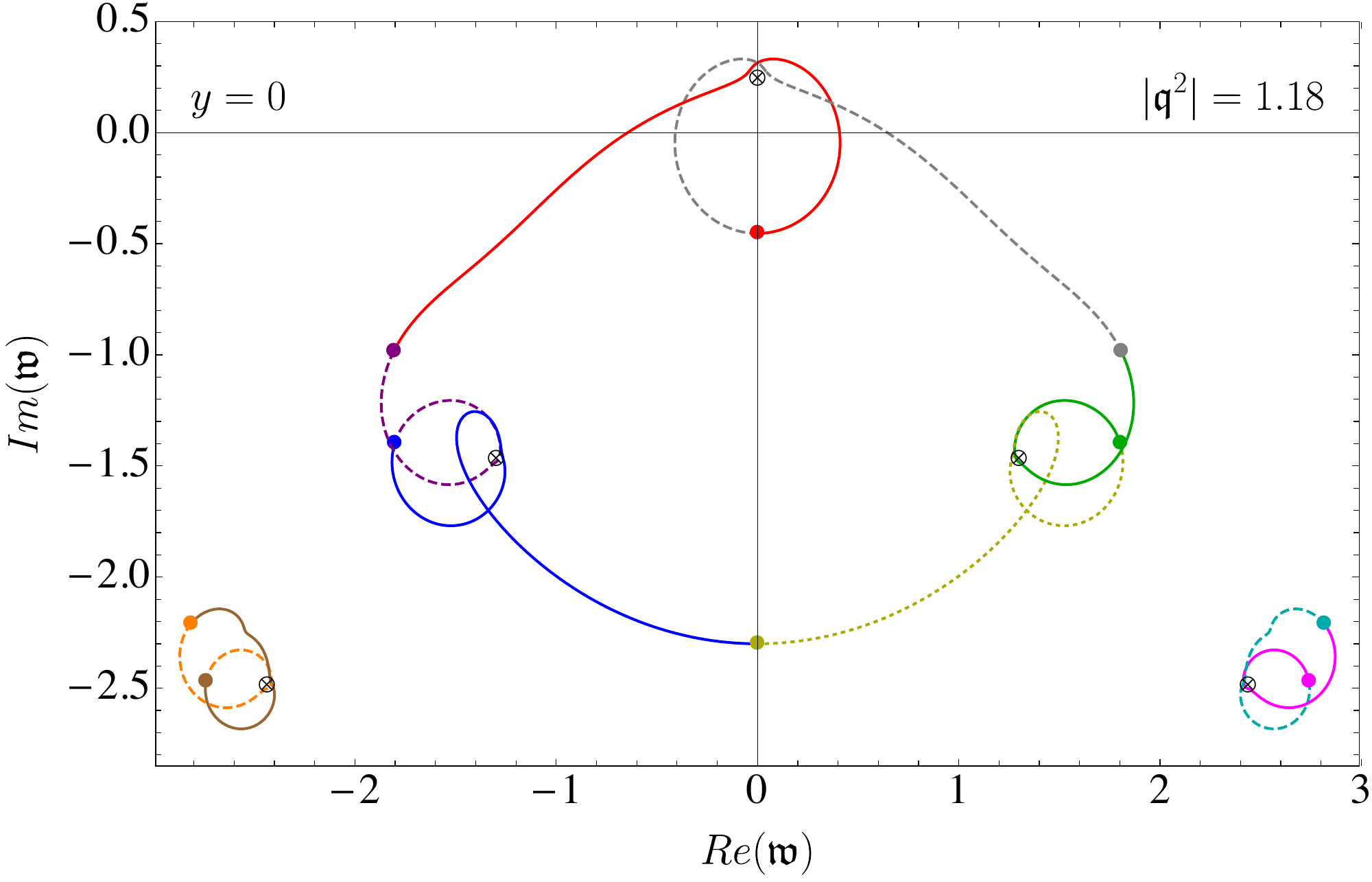}
    \includegraphics[width=0.47 \textwidth]{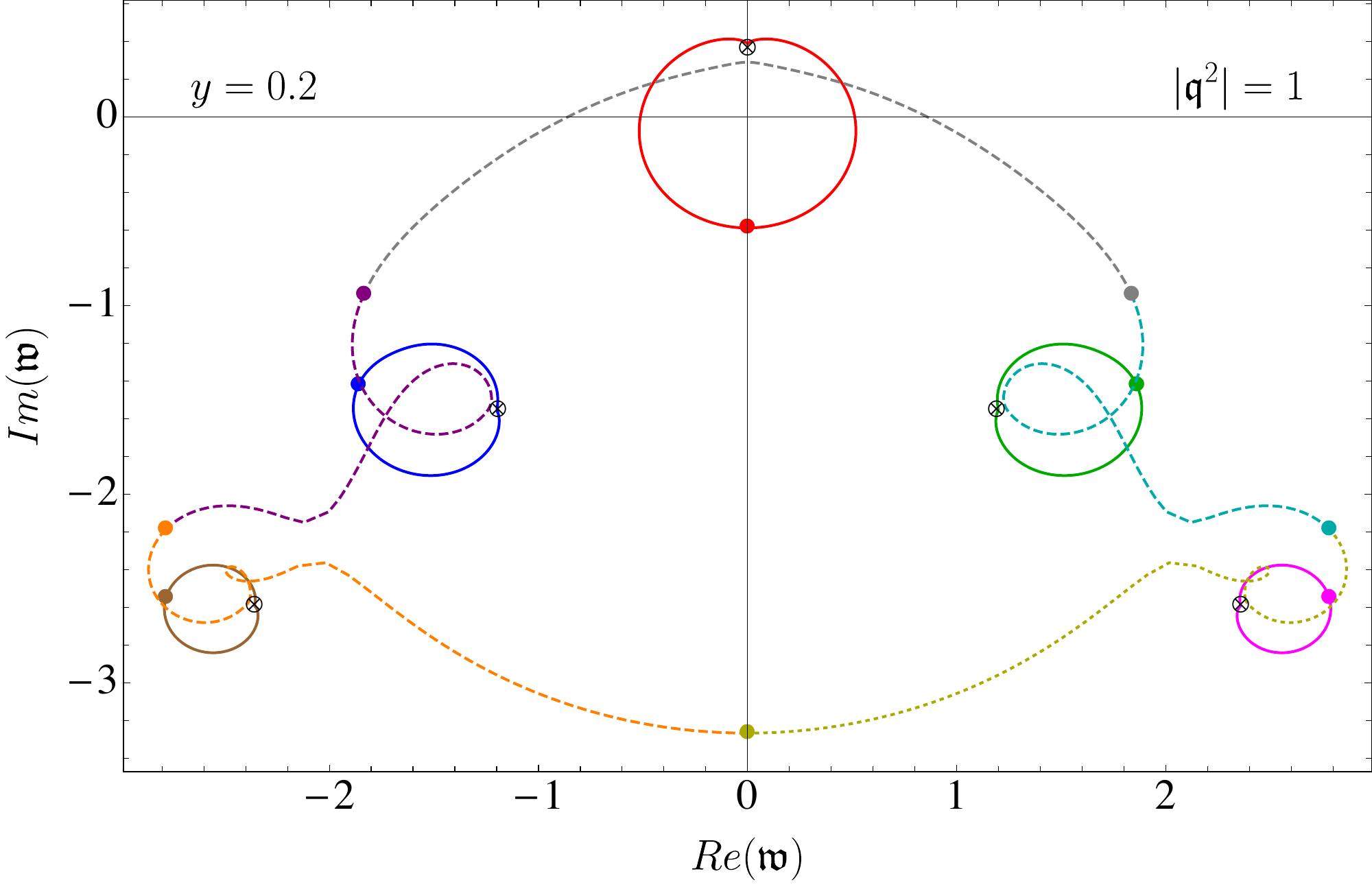}  
    \includegraphics[width=0.47 \textwidth]{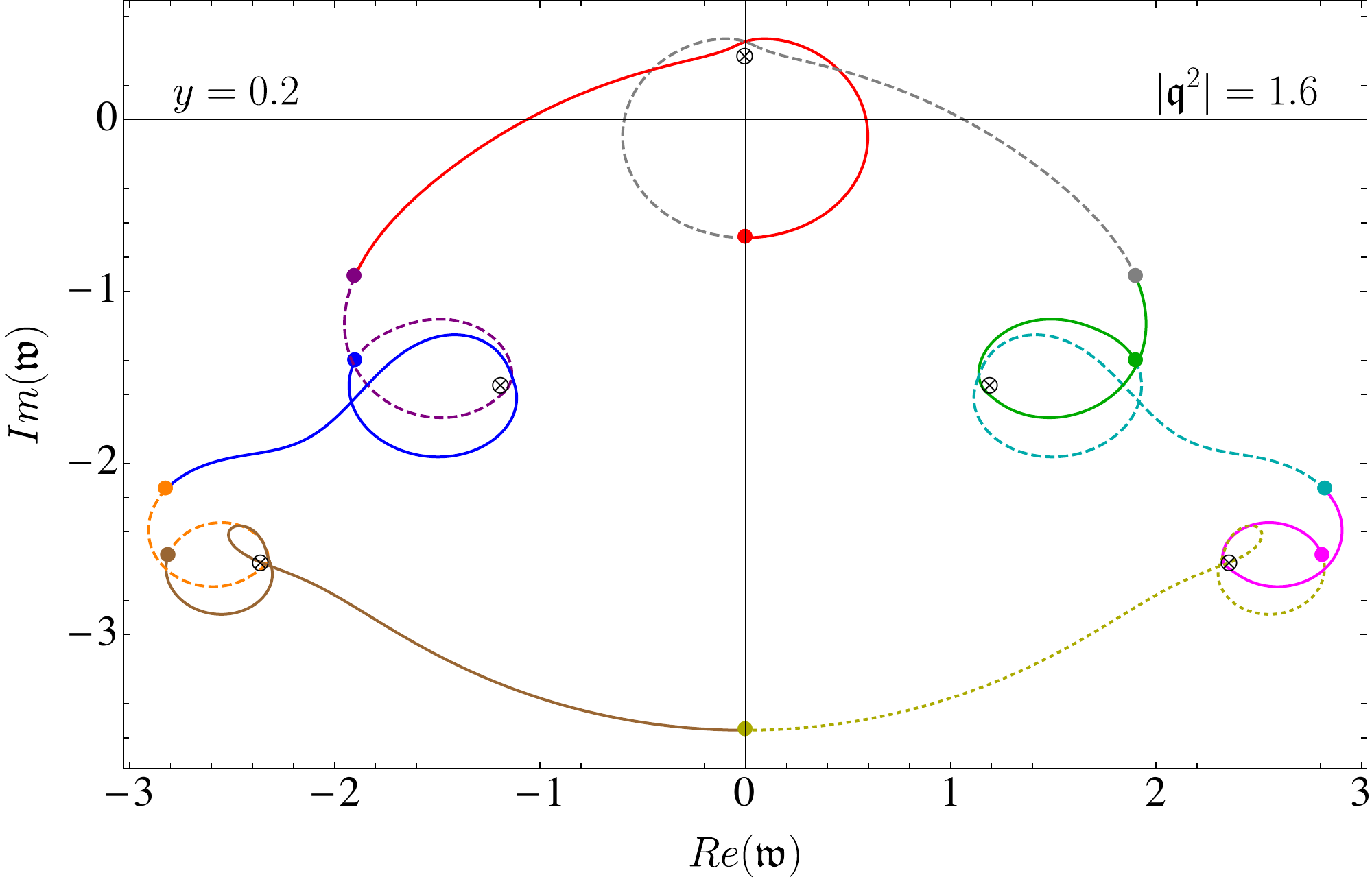}
    \caption{The trajectory of the lowest QNMs in shear channel for various values of the amplitude of the complex momenta square  $|\mathfrak{q}^2|$ for $y=0,~ 0.2$. We use the same colors as Figure \ref{fig:qnm-shear-y-1} to keep track of the shear modes and transverse gauge field modes in the coupled system. The joining happens at $\mathfrak{w}=0.244 i$ and $\mathfrak{w}=0.370 i$ respectively.}
    \label{fig:qnm-shear-y-0}
\end{figure}

There are couple of intriguing regimes of the $y$ parameter that we will investigate separately in the following. 
Let's start by recomputing the modes for AdS-Schwarzschild black brane  corresponding to $y=1$. In this case we can compare our results with the one which is already known \cite{Grozdanov:2019uhi}. We also introduce our conventions for the presentations of the trajectories of the modes. 
In the first row of Figure \ref{fig:qnm-shear-y-1}, we show the trajectories of the lowest  modes in spin-1 sector  as functions of complex momenta for $y=1$. In this case, for real momenta the QNM frequencies are two Christmas trees including  one for the  shear modes and one for the transverse  gauge modes  \cite{Kovtun:2005ev}. Interestingly, the first level-crossing occurs between the latter  modes at  $|\mathfrak{q}^2|=0.479$, while the level-crossing between the  hydro and first shear non-hydro mode occurs at $|\mathfrak{q}^2|=2.224$ which are in agreement with \cite{Grozdanov:2019uhi}. Therefore, by definition the radius of convergence of the hydrodynamic is $|\mathfrak{q}_*^2|=2.224$. Also note that for real momenta the non-hydro modes from two channels are always in pair and in the regime we are interested in, they are  such that the one from shear perturbation is less damped compare to the one from transverse gauge field perturbation. This will be the case  all the way to $y=0$. In all plots presented in this section, we use the same colors for the modes and their trajectories. The corresponding modes, for the real momenta are shown by dots and the positions of mode collisions are shown by crossed-circles.

While close to $y=1$ the convergence radius of the hydrodynamic series is due to a level-crossing between the hydro mode and the lowest non-hydro shear mode, for $0.596<y<0.89$ it's not the case anymore. The level-crossing will be replaced by a collision and  the hydro-mode will still have a closed trajectory.\footnote{By collision of the modes, we mean they swap part of their trajectories such that they still have disjoint trajectories.} This is demonstrated in the second row of the Figure \ref{fig:qnm-shear-y-1} for $y=0.8$. For this background the radius of convergence of the hydrodynamic series is $|\mathfrak{q}_*^2|=2.553$ which is associated with a mode collision  at $\mathfrak{q}_*^2=2.10475 \pm 1.44493 i$ and $\mathfrak{w}=\pm 1.48974\, -1.06122 i$. 

As we show in Figure \ref{fig:qnm-shear-y-k} the radius of convergence has a maximum value at $y=0.596$ and therefore one may expect that around this point something interesting could happen.  At $y=0.596$ the convergence radius of the hydrodynamic series is $|\mathfrak{q}_*^2|=2.854$ where two lines are crossed. Associated to this point, there is a collision between the hydro mode and  non-hydro shear modes at $\mathfrak{q}_*^2=2.28767 \pm 1.70642 i$ and   $\mathfrak{w}=\pm 1.5247 -1.0434 i$ as well as a level-crossing with  transverse gauge field non-hydro mode at $\mathfrak{q}_*^2=-2.854$ and $\mathfrak{w}=  0.86 i$.
The most important point here is that the  convergence radius of hydrodynamic series is found due to multi-phenomena: a level-crossing and a collision between the QNMs simultaneously.

Finally, in Figure \ref{fig:qnm-shear-y-0} we show  the trajectory of the least damped  modes in spin-1 sector  as functions of complex momenta at the critical point of the second order phase transition $y=0$ in the first row and $y=0.2$ in the second row. Again, the first level-crossing happens between the lowest non-hydro transverse gauge field modes.  This is always true for any value of $0\leq y \leq 1$. The radius of convergence of the hydrodynamic series is $|\mathfrak{q}_*^2|=\frac{9}{8}$ for $y=0$ and $|\mathfrak{q}_*^2|=1.47$ for $y=0.2$. Note that in this regime there are always infinite number of level-crossing between the shear and gauge field modes at the $|\mathfrak{q}_*^2|$, since the linearized equations  are the same at this critical value of the momenta, see the Equation \eqref{eq:master-eqs}. 
The radius of convergence of the hydrodynamic series at the critical point, given by $y=0$, is almost half of its counterpart $\mathcal{N}=4$ SYM without chemical potential, corresponding to $y=1$. 

Last but not least, in Figure \ref{fig:qnm-shear-y-0} there is a second damping mode which can be seen in small $y$ cases, as one of the lowest QNMs.
This observation may give us a hint about the QNM structure in the whole range of parameter space, in particular for $y=1$. As we already discussed at the critical value of momentum given in Equation \eqref{qc-analytic} two equations of motion in spin-1 sector coincide with each other. In other words, for each shear mode there should be a cousin in the transverse gauge field mode. On the other hand, due to the symmetries of the background and QNM equations in each sector, either the modes are purely imaginary or they are in pair as $\omega_{\pm}=\pm \omega_R- i \omega_I$. Having said that, we can conclude that there should be odd numbers of purely imaginary modes in the transverse gauge field channel to accomplish the fact that there are odd numbers of modes in the shear channel (one hydro and infinite number of pair non-hydro modes). Based on this argument we infer that in AdS-Schwarzschild black brane geometry the transverse perturbation of a gauge field should have at least one purely imaginary QNM which has not been addressed in the literature to our best knowledge. 

To summarize our results, for the convergence radius of the hydrodynamic series in shear channel we would like to point out that there is a competition  between the lowest non-hydro modes in the shear channel  and the ones in the transverse channel to join to the hydro-mode. In the range $0\leq y< 0.596$ which ends to the phase transition point the latter is responsible, while in $0.596<y\leq1$ the former takes the main role to specify the radius of convergence of hydrodynamic series. 

\section{Critical exponent}\label{sec:critical}

As mentioned before, the  theory we study in this paper enjoys a critical point at ${\mu}={\pi T}/{\sqrt{2}}$. The behavior of different observables near this critical point has been studied in \cite{DeWolfe:2011ts, Finazzo:2016psx, Ebrahim:2017gvk, Ebrahim:2018uky, Amrahi:2020jqg, Ebrahim:2020qif, Amrahi:2021lgh}. In all cases it was shown that the critical exponent is $\theta=\frac{1}{2}$.  All quantities  remain finite (except the diffusion constant which vanishes), while their slopes diverge at the critical point. 
By using the chain rule, it is easy to show that if a quantity $\mathcal{P}$ has the following expression  near the critical point in terms of the  dimensionless parameter $y=\sqrt{1-\frac{2 \mu^2}{\pi^2 T^2}}$ 
\be
\mathcal{P}- \mathcal{P}_c\sim y^m,
\ee
then, the behaviour of the associated quantity near critical point at either fixed temperature or fixed chemical potential will be given by,
\bea
&&\mathcal{P} - \mathcal{P}_c \sim |\mu - \mu_c|^{m/2}, \qquad \qquad T=\textrm{fixed},\\
&&\mathcal{P} - \mathcal{P}_c \sim |T - T_c|^{m/2}, \qquad \qquad \mu=\textrm{fixed},
\eea
where $\mathcal{P}_c$ is the critical value of the quantity.
That means for $m<2$ there is a critical exponent $\theta=\frac{m}{2}$ in both schemes. 
In particular, if a quantity is linear in $y$ close to the critical point, then the corresponding  critical exponent will be $\theta=\frac{1}{2}$.

Here, we would like to study the behavior of transport coefficients 
as well as the radius of convergence of the hydrodynamic series $|\mathfrak{q}_*^2|$ near the critical point and we choose to present the results for fixed temperature.  By using Equations \eqref{eq446}, \eqref{eq:theta1}, \eqref{eq:qc-spin-2}, \eqref{qc-analytic}, \eqref{eq:qc-spin-1} and expanding close to the critical point up to the linear term in $y$ we reach the following expressions
\begin{align}\label{eq:near-critical}
&\eta  \simeq  \frac{\pi  N_c^2 T_c^3}{8} \left(\frac{9}{8}+\frac{3}{8}y \right), \hspace{100pt}
\kappa  \simeq\frac{N_c^2 T_c^2}{8}(\frac{3}{4}+\frac{1}{2}y),\nn\\
& \tau_{\pi} \simeq\frac{2-\log (2)}{2 \pi  T_c} \left(\frac{4 \log (2)-4}{3 \log (2)-6}+\frac{4 \log (2)-10}{9 \log (2)-18}y\right), \hspace{20pt} \theta_1\simeq\frac{N_c^2T_c}{32\pi}y,\nn\\
& \lambda_{17}^{(3)}\simeq\frac{N_c^2 T_c\left(\pi ^2-12 (\log (2)-2) \log (2)\right) }{192 \pi } \times\nn\\
&\hspace{30pt}\left(\frac{9 \text{Li}_2\left(\frac{3}{4}\right)-6 (\log (2)-4) \log (2)}{\pi ^2-12 (\log (2)-2) \log (2)}+ \frac{3 \left(2 \text{Li}_2\left(\frac{3}{4}\right)-4-4 \log ^2(2)+8 \log (2)\right)}{\pi ^2-12 (\log (2)-2) \log (2)} y\right),\nn\\
& \lambda_1^{(3)} - \lambda_{16}^{(3)} \simeq \frac{N_c^2 T_c\left(\pi ^2-12 (\log (2)-4) \log (2)\right) }{192 \pi } \times\nn\\
&\hspace{30pt}\left(\frac{9 \text{Li}_2\left(\frac{3}{4}\right)-6 (\log (2)-8) \log (2)}{\pi ^2-12 (\log (2)-2) \log (2)}+ \frac{3 \left(2 \text{Li}_2\left(\frac{3}{4}\right)-4-2 \log ^2(2)+8 \log (2)\right)}{\pi ^2-12 (\log (2)-2) \log (2)} y\right),\nn\\
& |\mathfrak{q}_*^2|_\textrm{hydro spin-1}\simeq\frac{9}{8}+\frac{3}{2}y,\hspace{67pt}
|\mathfrak{q}_*^2|_\textrm{non-hydro spin-2}\simeq 1.505−0.634\, y,\nn\\
& -i\left(\mathfrak{w}_*\right)_\textrm{hydro spin-1}\simeq 0.244 + 0.531\, y, \hspace{20pt}
i\left(\mathfrak{w}_*\right)_\textrm{non-hydro spin-2}\simeq 1.637 + 0.685\, y.
\end{align}
Interestingly, all the quantities are linear in $y$ which means that they have  the same dynamical critical exponent $\theta =\frac{1}{2}$. 
It is noticed that in Equation \eqref{eq:near-critical} we organize the expressions such that the prefactors are the corresponding value for the AdS-Schwarzschild black brane\footnote{In the case of $|\mathfrak{q}_*^2|$ and $\mathfrak{w}_\star$ this form can not be applied.} which is a typical  critical exponent for the mean field theories \cite{Buchel:2010gd}, related to the Large $N_c$ limit \cite{Natsuume:2010bs}. It is intriguing to expect that other dynamical quantities such as higher transport coefficients may share the same critical behaviour close to the transition point.

Let us note that, although the background \eqref{metric} has linear expansion close to any value of $y_0$, namely
\begin{align}
g_{MN}&=g_{MN}^{(0)}+g_{MN}^{(1)}(y-y_0)+\mathcal{O}\left((y-y_0)^2\right),\nn\\
A_t&=A_t^{(0)}+A_t^{(1)}(y-y_0)+\mathcal{O}\left((y-y_0)^2\right),\nn\\
\phi&=\phi^{(0)}+\phi^{(1)}(y-y_0)+\mathcal{O}\left((y-y_0)^2\right),\nn    
\end{align}
the physical quantities associated to the boundary theory do not inherit this behaviour necessarily. 

\section{Conclusion}\label{sec:conclusion}
After reviewing general aspects of hydrodynamics as a gradient expansion in section \ref{sec:hydro},  Green's function and linear response theory in section \ref{sec:GreensF},  we focused on a specific example from section \ref{sec:1RCBH} onwards, namely the 4-dimensional $\mathcal{N}=4$ SYM theory at finite temperature and finite chemical potential which poses a second order phase transition at $\mu=\pi T/\sqrt{2}$. This theory is dual to a Einstein-Maxwell-dilaton  theory which is constructed by a consistent reduction and truncation of 10-dimensional type IIB supergravity on AdS$_5\times$S$^5$, keeping only the metric, one scalar field and one gauge field \cite{Gubser:1998jb, Behrndt:1998jd, Kraus:1998hv, Cai:1998ji, Cvetic:1999ne, Cvetic:1999rb}. The black hole solutions with planar horizon and with sourceless scalar field boundary condition in this theory are so-called 1RCBHs which are asymptotically AdS. In sections \ref{sec:hydro-1RCBH}-\ref{sec:critical} we computed the linear hydrodynamic transport coefficients up to the third order in spin-1 and spin-2 sectors as well as the radius of convergence of the hydrodynamic series in shear channel for arbitrary temperature and chemical potential. 

Surprisingly we found that close to the critical point of the phase transition all the quantities share the same critical exponent $\theta=\frac{1}{2}$ which might be related to the large $N_c$ limit of our setup \cite{Buchel:2010gd, Natsuume:2010bs}.\footnote{Although the Einstein-dilaton theory  with second order phase transition exhibit a universal critical exponent $\theta=\frac{2}{3}$ in thermodynamic quantities in various dimensions \cite{Ecker:2020gnw}.}
We speculate that the other hydrodynamic transport coefficients, including the non-linear ones, exhibit the same critical behaviour.  To  support our proposal let us recall a universal relation among the second order transport coefficients \cite{Erdmenger:2008rm, Haack:2008xx},
\be
4 \lambda_1^{(2)} +  \lambda_2^{(2)}=2 \eta \tau_\pi.\nn
\ee
which we expect to be hold in our setup too.

As one of our main results we found that  the only third order transport coefficient which appears in shear hydrodynamic dispersion relation  dies out at the critical point, $\theta_1=0$. This peculiar observation might be related to the symmetry enhancement of the underlying theory at the critical point, in a same spirit that the bulk viscosity vanishes due to conformal symmetry.  

Let us emphasize that  both analytical and  numerical studies in the spin-1 sector have been carried out by employing the master equations approach \cite{Kodama:2003jz, Kodama:2003kk, Jansen:2019wag} to solve the linearized equations of motion. To our best knowledge this is the first non-trivial example of Einstein-Maxwell-dilaton theory which benefits enormously from master equations formalism. We analyzed the QNM structure in the spin-1 sector with complex momentum to compute the convergence radius of the hydrodynamic series. We found that in different regime different mode collisions and/or level-crossing is responsible for radius of convergence. In a range of parameters which ends to the critical point ($0\leq y \leq 0.596$) we found a perfect agreement between our numerical results and the analytical findings which is a consequence of having the same linearized equations for both shear channel and transverse channel at special momentum. Therefore, in this regime the lowest level-crossing that fixes the hydrodynamics radius of convergence is between the shear hydro mode and the lowest transverse non-hydro modes. On top of that we found that the minimum radius of convergence occurs at the critical point.
On the other hand, in the second regime which ends to the zero chemical potential ($0.596\leq y \leq 1$) we found that a mode collision or a level-crossing between the shear hydro mode and lowest shear non-hydro modes lead to the radius of convergence of the hydrodynamic series. 
For completeness, we have also studied the spectrum of QNMs in spin-2 sector. Due to the fact that there is no  hydro mode in this sector, the radius of convergence is set by the collision of non-hydro modes and we have analyzed it in the entire range of the parameter space.

Now we would like to  address some generalizations of our results.
As we already mentioned, the only third order transport coefficient which shows up in shear dispersion relation, $\theta_1$, vanishes at the critical point.
This suggest that the full hydrodynamic expansion may contain less number of transport coefficients at the critical points, even at the higher order of expansion. In this manner, further investigations for other cases deserve more attention in future studies.
To complete our results, the next step is to study the spin-0 sector of the same model which includes three coupled linearized equations. With some efforts we found numerically that the other third order transport coefficient $\theta_2\equiv-(\lambda^{(3)}_3+\lambda^{(3)}_5+\lambda^{(3)}_6)$ which appears in the sound hydro mode dispersion relation \cite{Grozdanov:2015kqa} does not vanish at the critical point of the phase transition. It turns out that a complete investigation of this sector with complex momentum is much more involved and has some subtleties that we leave this study for future work. 

It could also be of interest to venture other  models  with second order phase transition, e.g. bottom-up holographic cases \cite{Hartnoll:2008kx, DeWolfe:2010he, Janik:2015iry, Janik:2016btb, Ecker:2020gnw}, to study the radius of convergence of the hydro-mode and to check our proposal of a relation between the symmetry enhancement of the underlying theory and vanishing higher order transport coefficients at the critical point.

Finally, it will be interesting to  consider the large but finite coupling impacts to our findings with an ultimate goal of interpolating between weak
and strong coupling results. This can be achieved for example by adding the Gauss-Bonnet term to the action. The QNMs and some of the transport coefficients of such a theory at zero chemical potential  have been already computed in \cite{Grozdanov:2016fkt}.

\section*{Acknowledgement}

We would like to thank Michal Heller, Romuald Janik, Jakub Jankowski, Andrzej Rostworowski, Michal Spalinski and Andrei Starinets for valuable discussions  and their comments on the first version of this manuscript. 
This work is partly supported by Guangdong Major Project of Basic and Applied Basic Research No. 2020B0301030008, the National Natural Science Foundation of China with Grant No.12035007, Science and Technology Program of Guangzhou No. 2019050001.

\begin{appendix}
	
\section{Spin-2 perturbation }\label{AppA}
Here we will address the details of calculations of section \ref{sec:spin-2}. We shall show that how to derive $\mathcal{H}(r)$ generally from variation of parameters method. This is a necessary step to obtain the transport coefficients.

Generally speaking, the Equation \eqref{eq23} is classified as the Heun differential equation \cite{Kristensson:2010GK} and it has four types of different singular points. To our purpose, only two of the singular points are important which are at the horizon and at boundary,
\begin{align}\label{eqApp:3}
	r_\star = 1,\,  0.
\end{align}
The singular points in the Heun equation are regular-singular points and near these points we can solve the differential equation using the Frobenius ansatz  
\begin{align}\label{eqApp:4}
	H_{xy}(r) \to (r-r_{\star})^{a} \widetilde{H}(r),
\end{align}
where $\widetilde{H}(r)$ is a regular function at $r_{\star}$. The index $"a"$ can be obtained by the regularity condition. 
By plugging the ansatz  \eqref{eqApp:4} in linearized equation of motion \eqref{eq23}
one can solve the corresponding indicial equation \cite{Kristensson:2010GK}.  Near our singular points \eqref{eqApp:3} the results are
	\begin{align}\label{eqApp:5}
	    &\widetilde{H}(r) \to (1-r^2)^{-i\frac{\mathfrak{w}}{2}} \widetilde{H}^{(h)}_1(r) + (1-r^2)^{i\frac{\mathfrak{w}}{2}}\widetilde{H}^{(h)}_2(r) + \cdots,\nn\\
	    &\widetilde{H}(r) \to \widetilde{H}^{(b)}_1 + \widetilde{H}^{(b)}_2 r^4 + \cdots.
	\end{align}
 Physical conditions dictate which term has to be picked up. For ingoing wave solution near the horizon we should take
\begin{align}\label{eqApp:6}
	H_{xy}(r) = (1-r)^{-i \frac{\mathfrak{w}}{2}} \mathcal{H}(r).
\end{align}
Likewise, the Dirichlet boundary condition rules that $\widetilde{H}^{(b)}_1(r\to 0) =0$. This boundary condition will give  the spectrum of QNMs \cite{Kovtun:2005ev}. It is worthwhile to mention that values of index $"a"$ depend on  the chosen coordinates. For example in the EF coordinates where $g^{tt}=0$, the indices are derived as
\begin{align}\label{eqApp:7}
    &a = 0, \,\,i \mathfrak{w}, \qquad \qquad \mbox{As $r \to $ 1},\nn\\
    &a=0,\,\,4 , \qquad \qquad \,\,\,\,\mbox{As $r \to $ 0}.
\end{align}
As we stated above, solving exactly the Equation \eqref{eq23} is a tedious job. But in the hydrodynamics  limit $(\mathfrak{w}, \mathfrak{q})\ll 1$, we can solve it by the method of variation of parameters  \cite{Grozdanov:2019uhi}.  
By plugging  the ansatz \eqref{eqApp:6} into the Equation \eqref{eq23} and using the  following expansion 
\begin{align}\label{eqApp:8}
    \mathcal{H}(r) = \sum_{n=0}^\infty \epsilon^n \mathcal{H}_n(r), \qquad (\mathfrak{w}, \mathfrak{q}) \to (\epsilon \mathfrak{w}, \epsilon \mathfrak{q}),
\end{align}
 for each $\mathcal{H}_n(r)$ the following equation would appear
\begin{align}\label{eqApp:9}
&\mathcal{H}''_n(r) - \frac{r^4(3-y) + 2r^2(1-y) + 3(1+y)}{r (1-r^2) (1+y+r^2(3-y))} \mathcal{H}'_n(r)\nn\\
& = -\frac{2 i r}{1-r^2}\mathfrak{w} \mathcal{H}'_{n-1}(r) + \mathcal{C}_1 \mathfrak{w} \mathcal{H}_{n-1}(r) + \left(\mathcal{C}_2 \mathfrak{w}^2  + \mathcal{C}_3 \mathfrak{q}^2 \right) \mathcal{H}_{n-2}(r).
\end{align}
The $\mathcal{C}_{1, 2, 3}$ are defined in below
\begin{align}\label{eqApp:10}
\mathcal{C}_1 & \equiv \frac{2 i (1+y)}{(1-r^2) (1+y + r^2 (3-y))},\nonumber\\
 \mathcal{C}_2& \equiv -\frac{r^4(3-y)^3 + 16 (1+y) + r^2 (3-y)^2 (5+y)}{(1-r^2) (3-y) (1+y + r^2 (3-y))^2},\nn\\
 \mathcal{C}_2& \equiv \frac{16}{(1-r^2) (3-y) (1+y + r^2 (3-y))}.
\end{align}
Homogeneous solutions for the Equation \eqref{eqApp:9} are given by $\mathcal{H}_0(r) = c_1 \mathfrak{h}_1(r) + c_2 \mathfrak{h}_2(r)$ where
\begin{align}
    \mathfrak{h}_{1}(r) = 1, \qquad \quad \mathfrak{h}_2(r) = \frac{1+y}{3-y}\log (1 + y +r^2 (3-y)) + \log (1-r^2).
\end{align}
Method of variation of parameters states that having the homogeneous solutions is enough to find the general solution of the Equation \eqref{eqApp:9} \begin{align}\label{eqApp:12}
    \mathcal{H}_n(r) = c_1 \mathfrak{h}_1(r) + c_2 \mathfrak{h}_2(r) + \mathfrak{h}_1(r) \int_r^1 du \,\, \frac{\mathfrak{h}_2(u) \mathcal{F}_n(u)}{W(\mathfrak{h}_1, \mathfrak{h}_2)} - \mathfrak{h}_2(r) \int_r^1 du \,\, \frac{\mathfrak{h}_1(u) \mathcal{F}_n(u)}{W(\mathfrak{h}_1, \mathfrak{h}_2)}.
\end{align}
Here, $W(\mathfrak{h}_1, \mathfrak{h}_2)$ is the Wronskian of the two homogeneous solutions and $\mathcal{F}_n(u)$ is  right hand side of the Equation \eqref{eqApp:9}. The boundary condition at $r=1$ demands that the integration constants should vanish, $c_1=0, c_2=0$, for $n> 1$. 
This recursive equation means that having $\mathcal{H}_0(r)$ and $\mathcal{H}_1$ one can find $\mathcal{H}_2$ and so on. In the following we show the results  up to the third order
\begin{align}\label{eqd3}
\mathcal{H}_0(r)& =1, \\
\mathcal{H}_1(r) &= -i \frac{\mathfrak{w}}{2} \frac{1+y}{3-y} \log \frac{1+ y + r^2(3-y)}{4},\nn\\
\mathcal{H}_2(r) &=\frac{4\left(\mathfrak{w}^2 -\mathfrak{q}^2\right)}{(3-y)^2} \log \frac{1+ y + r^2(3-y)}{4} \nonumber\\
&\quad + \frac{\mathfrak{w}^2 (1+y)}{8 (3-y)} \left\{\log^2 \bigg(\frac{1+ y + r^2(3-y)}{4}\bigg) - \frac{8}{1+y} \text{Li}_2\left(\frac{(1-r^2)(3-y)}{4}\right)\right\},\nonumber\\
\mathcal{H}_3(r) &=\frac{i\mathfrak{w}}{{48 (y-3)^2}} \Bigg\{ 24 \text{Li}_2\left(\frac{(r^2-1) (y-3)}{4} \right) \left(8 (\mathfrak{q}^2-\mathfrak{w}^2 )+(y+1) \mathfrak{w} ^2 \log \left(\frac{r^2 (3-y)+y+1}{4} \right)\right)\nn\\
&\hspace{80pt}+\mathfrak{w} ^2 \bigg((y+1) \bigg((7-y) \log^3 \left(\frac{r^2 (3-y)+y+1}{4}\right)\nn\\
&\hspace{80pt}-48 \text{Li}_3\left(\frac{r^2 (3-y)+y-4}{r^2 (3-y)+y+1}\right)\bigg)-192 \text{Li}_3\left(\frac{(r^2-1) (y-3)}{4}\right)\bigg)\Bigg\}\nn.
\end{align}

\end{appendix}

\bibliographystyle{fullsort}
\bibliography{references}

\end{document}